%% file: main.tex
\documentclass[11pt]{article}
\usepackage{fullpage}

\input{macro}

 \title{\bf \textit{Who Are We Missing?}: A Principled Approach to Characterizing the Underrepresented Population}
  \author{Harsh Parikh\\
    Department of Biostatistics, Johns Hopkins University\\\\
    Rachael K. Ross \\
    Department of Epidemiology, Columbia University \\\\
    Elizabeth Stuart\\
    Department of Biostatistics, Johns Hopkins University\\\\
    Kara E. Rudolph\\
    Department of Epidemiology, Columbia University}

\date{}

\begin{document}

\maketitle

\input{abstract}
\input{introduction_new}
\input{literature}
\input{prelim}
\input{goalpost}
\input{estimation}
\input{moud_trial}
\input{conclude}
\bibliographystyle{apalike}
\bibliography{biblio}
\pagebreak
\begin{appendices}
    \input{appendix_proof_prelim}
    \input{appendix_alg_root}
    \input{appendix_theory}
    \input{results}
    \input{appendix_table}
    \input{moud_discuss_assume}
\end{appendices}

\end{document}

%% file: macro.tex
\usepackage{amsmath}
\usepackage{graphicx}
\usepackage{amssymb}
\usepackage{amsthm}
\usepackage{enumitem}
\usepackage{bbm}
\usepackage[english]{babel}
\usepackage[titletoc,title]{appendix}
\usepackage{natbib}

\theoremstyle{remark}
\newtheorem{remark}{Remark}
\usepackage{algorithm}
\usepackage{algpseudocode}
\usepackage{booktabs}
\usepackage{hyperref}
\hypersetup{
    colorlinks=true,
    linkcolor=red,
    citecolor=blue,
    filecolor=magenta,      
    urlcolor=cyan}

\newcommand{\bX}{\mathbf{X}}
\newcommand{\bC}{\mathbf{B}}

\newcommand{\bx}{\mathbf{x}}
\newcommand{\bc}{\mathbf{b}}
\newcommand{\ba}{\mathbf{a}}

\newcommand{\E}{\mathbb{E}}
\newcommand{\bV}{\mathbf{V}}

\newcommand{\bv}{\mathbf{v}}

\newcommand{\setSn}{\mathcal{D}_n}

\newcommand{\setAn}{\mathcal{A}_n}

\newcommand{\muXt}{\mu(\bX,t)}
\newcommand{\muxt}{\mu(\bx,t)}
\newcommand{\muX}[1]{\mu(\bX,{#1})}
\newcommand{\p}{\pi}
\newcommand{\pX}{\p(\bX)}
\newcommand{\px}{\p(\bx)}
\newcommand{\el}{\ell}
\newcommand{\lX}{\el(\bX)}
\newcommand{\lx}{\el(\bx)}

\newcommand{\atte}{{\tau}_0}

\newcommand{\attehat}{\widehat{\tau}_0}

\newcommand{\phat}{\widehat{\p}}

\newcommand{\lhatXi}{\widehat{\el}(\bX_i)}

\newcommand{\yhat}{\widehat{Y}}
\newcommand{\ehat}{\widehat{e}}

%% file: abstract.tex
\begin{abstract}
Randomized controlled trials (RCTs) serve as the cornerstone for understanding causal effects, yet extending inferences to target populations presents challenges due to effect heterogeneity and underrepresentation. Our paper addresses the critical issue of identifying and characterizing underrepresented subgroups in RCTs, proposing a novel framework for refining target populations to improve generalizability. We introduce an optimization-based approach, Rashomon Set of Optimal Trees (ROOT), to characterize underrepresented groups. ROOT optimizes the target subpopulation distribution by minimizing the variance of the target average treatment effect estimate, ensuring more precise treatment effect estimations. Notably, ROOT generates interpretable characteristics of the underrepresented population, aiding researchers in effective communication. Our approach demonstrates improved precision and interpretability compared to alternatives, as illustrated with synthetic data experiments. We apply our methodology to extend inferences from the Starting Treatment with Agonist Replacement Therapies (START) trial -- investigating the effectiveness of medication for opioid use disorder -- to the real-world population represented by the Treatment Episode Dataset: Admissions (TEDS-A). By refining target populations using ROOT, our framework offers a systematic approach to enhance decision-making accuracy and inform future trials in diverse populations. 
\end{abstract}


%% file: introduction_new.tex
\section{Introduction}
Randomized controlled trials (RCTs) are the gold standard for estimating treatment effects due to the internal validity they achieve by randomly assigning participants to treatment arms \citep{ding2023first}. However, the characteristics of the trial cohort often differ from the target population about whom one would like to draw inferences \citep{degtiar2023review}. Further, if the misaligned population characteristics between the trial and the target population are also effect modifiers, this can challenge the generalizability of trial findings \citep{degtiar2023review, rudolph2023efficiently}. 

Existing approaches generalizing trial inferences adjust for differences in the distribution of population characteristics between the trial cohort and the target population \citep{degtiar2023review}. However, many of these estimators lack precision while estimating effects over subgroups insufficiently represented in the trial cohort. This can lead to treatment decisions based on limited data and strong assumptions. Despite this concern, there is limited focus on identifying regions of the target population where inferences are most reliable and, crucially, areas where the trial provides little reliable information. Existing methods that focus on characterizing underrepresented subgroups do not account for outcome variability and only account for sample selection probability \citep{stuart2011, tipton2014, ding2016, tipton2017, cahan2017computer, greenhouse2008}. Further, they rely on summary measures such as selection scores that are inherently hard to interpret.

\paragraph{Contributions.} In this paper, we aim to interpretably characterize subgroups for which precise estimation of treatment effects is particularly challenging. These subgroups often occupy regions of the covariate space with heterogeneous treatment effects and insufficient representation in the trial data. We refer to these subgroups as ``insufficiently represented'' or ``underrepresented'' populations, due to their lack of data support in the trial, which hampers precise treatment effect estimation. Our approach involves identifying trial units that significantly contribute to the variance in estimating the target average treatment effect (TATE). We also redefine the estimand to focus on inferences within sufficiently represented subpopulations. This introduces two tensions: (i) a conceptual tension between  confidence in the inferences versus the extent of their generalizability, and (ii) a computational tension between dropping units to minimize variance versus maximizing sample size for statistical power.

We formalize this computational tension as a functional optimization problem and introduce an interpretable, non-parametric approach for functional optimization called the Rashomon Set of Optimal Trees (ROOT)\footnote{Rashomon is a classic Japanese movie in which an event is given more than one and often a competing interpretation by the individuals involved. It provides different and sometimes contradictory points of view on the same event. Here, each of the trees in the set provides almost equally good but the different characterization of underrepresented subpopulations -- thus we refer to this set as a Rashomon Set. Analogously, there is also literature in Interpretable Machine Learning that focuses on understanding the properties and behavior of almost equally good but different models \citep{breiman2001statistical, semenova2022existence}.}. By interpretable, we mean that researchers can effectively communicate the characteristics of insufficiently represented groups -- this helps address the conceptual tension by characterizing the population for whom the inferences can be made with high confidence. Given the trial and target data, ROOT identifies trial subpopulations that contribute significantly to the variance of the target treatment effect estimate. This high variance arises from a lack of balance in potential effect modifiers between the trial and target samples. In other words, ROOT identifies subpopulations with a small estimated probability of being in the trial population conditional on the effect modifiers—i.e., those units that may exhibit practical violations of positivity (discussed technically in Section~\ref{sec: prelim}). ROOT uses a tree-sampling-based algorithm with an explore-exploit search strategy to find the optimal solution to the functional optimization problem, accounting for potential effect heterogeneity and differences in characteristics between trial and target populations. We construct a set of trees with near-optimal objective values, recognizing the existence of multiple equally good models with different representations and interpretations, called a Rashomon set \citep{xin2022exploring}. 

We apply our approach to investigate the research-to-practice gap in studies of medication for opioid use disorder (MOUD), where discrepancies between treatment effectiveness measured in trials and real-world care settings have been noted. Recent work has highlighted that this discrepancy can be due to underrepresentation in trial cohorts \citep{rudolph2022under}. We address this by identifying underrepresented groups in an RCT of MOUD and extending inferences from that trial to a refined target population that excludes these underrepresented groups. Specifically, we consider the Starting Treatment with Agonist Replacement Therapies (START) MOUD trial \citep{saxon2013buprenorphine, hser2014treatment}, which compares methadone with buprenorphine, and the real-world population of individuals for whom MOUD is part of their treatment plan, using the Treatment Episode Dataset: Admissions (TEDSA) \citep{abuse2020treatment}. This dataset includes individuals treated for substance use disorders at programs across the United States that receive public funding.

\paragraph{Paper Organization.} We first summarize the relevant literature in Section~\ref{sec: literature}. In Section~\ref{sec: prelim}, we introduce our setup and assumptions. We propose our framework and the ROOT approach in Sections~\ref{sec: framework} and \ref{sec: method}. We then proceed to demonstrate the effectiveness of our approach through synthetic data experiments in Section~\ref{sec: synth_exp}. Lastly, in Section~\ref{sec: moud}, we discuss the application of our methodology to the MOUD case study. We end with a discussion of the limitations of our approach, posing open questions and future directions (Section~\ref{sec: conclude}).

%% file: literature.tex
\section{Relevant Literature}\label{sec: literature}
The literature on generalizability and transportability focuses on extending inferences from RCTs to a target population of interest. Generalizability pertains to situations where the trial cohort is a subset of the target population \citep{degtiar2023review}. Transportability deals with cases where the trial cohort is (at least) partially external to the target population \citep{pearl2011, degtiar2023review}. The success of generalizing or transporting causal effects might be compromised due to differences in individuals' observed and unobserved characteristics, treatment administration mechanisms, the mechanisms by which treatment affects the outcome, and outcome measurements between the trial and target populations \citep{rothwell2005, green2006, dekkers2010, burchett2011, degtiar2023review}. Our paper focuses on the differences in the pre-treatment characteristics between the trial and the target populations; the other factors, though important, are beyond the scope of this paper.

A challenge to generalizing or transporting effect estimates from trials to target populations is a potential lack of representativeness of the trial cohort. Current approaches in the literature diagnose this difference between the trial and target samples via distance measures. While some of these methods directly compare covariate distributions \citep{cahan2017computer, greenhouse2008, tipton2017}, other approaches compare the distribution of distance metrics that are functions of covariates, such as comparing the distribution of selection scores (estimated probabilities of participation in the trial as a function of covariates) across the trial and target samples \citep{stuart2011, tipton2014, ding2016, tipton2017}. 

Methods that directly compare covariate distributions include, for example, that of \cite{cahan2017computer}, which compares covariate averages across studies as a ratio of the mean or median values in the trial and target samples, and summarizes the discrepancies into a generalization score. Alternatively, instead of comparing averages, \cite{greenhouse2008generalizing} tests for the distributional discrepancy across each covariate, one at a time, adjusting for multiple hypothesis testing. 
Analogously, absolute standardized mean differences for each covariate are commonly used to measure overlap \citep{tipton2017}. While these covariate-based comparisons are common and can be helpful, they are limited in that they compare only the marginal distributions of each covariate. Moreover, in high-dimensional scenarios, this approach becomes impractical, potentially resulting in false positives or is more likely to identify misalignment on covariates not impacting the treatment effect. 


Methods that compare functions of covariates, typically involve comparing selection scores between trial and target samples, assessing the likelihood of an individual with specific covariate values being in the trial sample compared to the target population \citep{stuart2011, tipton2014}. A selection score serves as a unidimensional summary of covariates, capturing them as the estimated probability of being in the trial conditional on the covariates. Significant overlap in the selection score distributions of trial and target units suggests an effective representation of the target population by the trial cohort. This process parallels assessment of propensity score overlap between treatment groups in observational studies. Various distance measures, including Levy distance, Kolmogorov-Smirnov distance, Q-Q plots, and the overlapping coefficient, can be employed to evaluate and characterize the extent of overlap between the trial sample and target population \citep{ding2016, tipton2017}. To assess the extent of generalizability, it is possible to examine the overlap in selection score distributions across the trial and target samples. For instance, the analysis can be narrowed down to the target sample, specifically those with selection scores falling within the 5th and 95th percentiles of the trial sample's selection scores. \citep{tipton2014, tipton2017, sturmer2021propensity}. However, such trimming approaches lack a principled approach concerning a specific quantity of interest. Further, the above approaches are limited in that they do not provide an interpretable characterization of the underrepresented or refined target population, and may be highly sensitive to the choice of selection score model. 

Another key limitation of the above existing approaches is that 
they do not distinguish variables that matter for generalization -- the effect moderators -- from those that do not. Consequently,  
current methods may erroneously label certain subgroups as ones where trial inferences cannot be reliably extended. 

As mentioned above, issues around potential lack of representation in trial data has parallels with the challenge of overlap between treatment groups in observational studies. In an observational study, the imputation of missing potential outcomes relies on overlap across the treatment arms, and a lack of sufficient overlap across treatment arms also leads to a lack of precision in causal effect estimates. \cite{crump2009dealing} and \cite{li2019addressing} present a principled approach to characterize the population with appropriate overlap in observational studies, enabling the estimation of average treatment effects with high precision. These approaches involve redefining the estimand as the weighted average treatment effect where the weights are a function of pretreatment covariates. In a specific case with binary weights, these weights determine which subpopulation is included in the analyses and which subgroups are pruned.

Our approach extends the reweighted estimand concept, adapting it to the challenge of generalizing/transporting causal effects and characterizing the target population for which precise generalization is feasible. Unlike existing methods, we learn weights as a direct function of the covariates to ensure interpretability. We elaborate on our setup and methodology in the remainder of the paper.

%% file: prelim.tex
\section{Preliminaries}\label{sec: prelim}

\input{prelim/setup}
\input{prelim/assumptions}
\input{prelim/identification}
\input{prelim/estimate}

%% file: prelim/setup.tex
\paragraph{Setup}
In this study, we are interested in generalizing the causal effect of treatment ($T$) on outcome ($Y$) from an RCT to a target population. Our dataset comprises of $n$ individuals $\setSn = {1, \ldots, n}$. Corresponding to each unit $i \in \setSn$ and binary treatment choice, $1$ and $0$, there are two potential outcomes, represented as $Y_i(1)$ and $Y_i(0)$, respectively. We define the individual treatment effect, $\tau_i$, as the difference between these potential outcomes for person $i$, $\tau_i := Y_i(1) - Y_i(0)$. Each unit $i \in \setSn$ is categorized as either belonging to the RCT ($S_i = 1$) or the target population data ($S_i = 0$). In the RCT, we observe a $p$-dimensional vector of pretreatment covariates ($\bX_i = (X_{1,i}, \ldots, X_{p,i}$)), a treatment indicator ($T_i$), and one of the two potential outcomes for each person, depending on their treatment assignment: $Y_i = T_i Y_i(1) + (1 - T_i) Y_i(0)$. However, for units in the target population ($S_i = 0$), we observe only their covariates $\bX_i$. To simplify our discussion, we assume Bernoulli randomization in the RCT. For clarity, we denote $n_1$ as the number of units in the RCT ($S = 1$) and $n_0$ as the number of units in the target sample ($S = 0$), such that $n = n_1 + n_0$. Our primary estimands of interest are the target average treatment effect (TATE): $\atte := \E[\tau_i | S_i = 0]$, and the sample target average treatment effect (STATE): $\atte^{\setSn}:= \frac{1}{n_0} \sum_i (1 - S_i) \tau_i$. Here, STATE is a finite sample equivalent version of the TATE.

%% file: prelim/assumptions.tex
\paragraph{Assumptions.}
Now, we discuss the assumptions common in the literature regarding the identification of TATE and STATE:
\begin{enumerate}[label=A.\arabic*.]
    \item \textit{(Positivity)} For all $\bx$, the probability of participating in the trial is bounded away from 0 and 1: $0 < P( S=1 \mid \bX=\bx ) < 1$, and within the trial, the propensity of receiving treatment is bounded away from 0 and 1: $0 < P(T=1 \mid \bX=\bx, S=1) < 1$. \label{a: positive}

    \item \textit{(Internal Validity)} In the  RCT, treatment assignment is independent of the potential outcomes: $(Y(1), Y(0)) \perp T \mid \bX, S=1$. \label{a: internal_val}

    \item \textit{(External Validity)} The potential outcomes are independent of the trial participation conditional on the pre-treatment covariates $\bX$: $(Y(1), Y(0)) \perp S \mid \bX$. \label{a: external_val}

    \item \textit{(Stable Unit Treatment Value)} For unit $i$ in the RCT, i.e. $S_i=1$ and any pair of treatment assignments $\{T_1,\dots,T_{n_1}\}$ and $\{T'_1,\dots,T'_{n_1}\}$, if $T_i = T'_i$ then $Y_i(T_i) = Y_i(T'_i)$.  Further, for any pair of sample assignments $\{S_1,\dots,S_n\}$ and $\{S'_1,\dots,S'_n\}$, if $S_i = S'_i$ then $Y_i(t, S_i) = Y_i(t, S'_i)$ for all $t\in\{0,1\}$.\label{a: sutva}
\end{enumerate}

\paragraph{Discussion of Assumptions} These assumptions are also discussed in \citet{stuart2011} and \citet{tipton2013}. Assumption \ref{a: positive} establishes the support of the population distribution, ensuring that the probability of observing any point within the support is non-zero. This standard assumption is fundamental in causal inference, as it guarantees the existence of a comparable unit to estimate the missing counterfactual. While the second part of the assumption -- $0 < P(T=1 \mid \bX, S=1) < 1$ -- is guaranteed due to the design of the experiment; people are not enrolled in a trial if they can't have either treatment. However, the first part -- $0 < P(S=1 \mid \bX) < 1$ --  can be challenged when units with specific covariate values are systematically excluded from the RCT or never receive treatment. In our work, we assume that structural positivity holds i.e., $P(S=1 \mid \bX=\bx) \neq 0$, however, we focus on scenarios with practical violations of positivity (as defined in \citet{petersen2012diagnosing}) where certain subpopulations are insufficiently represented in the trial cohort.
Assumption \ref{a: internal_val} is generally upheld in well-implemented trials. 
Assumption \ref{a: external_val} implies that there are no unobserved factors differentiating units in the trial from the population.
Under this assumption, $P(Y(t) | \bX = \bx, S = 1) = P(Y(t) | \bX = \bx, S = 0)$. This assumption is also referred to as ``S-admissibility'' \citep{egami2023elements, pearl2011}.

%% file: prelim/identification.tex
\paragraph{Identification.}
We define a few more quantities for notational convenience and parsimony: (i) expected outcome, $\muXt:= \E( Y \mid \bX, T=t, S=1)$, (ii) selection score, $\pX:= P( S=1 \mid \bX )$, (iii) selection ratio, $\lX:= \frac{\pX}{1-\pX}$, (iv)$\p:= P( S=1 )$, and (v) trial propensity score, $e(\bX):= P( T=1 \mid \bX, S=1 )$.
Then, given assumptions \ref{a: positive}--\ref{a: external_val}, TATE $\atte$ is identified by $\theta_0 := \E_{\bX \mid S=0}(\muX{1} - \muX{0}) \mathrm{, and}$ $\theta_1 := \frac{\p}{1-\p} \E_{\bX \mid S=1} \left( \frac{\muX{1}-\muX{0}}{\lX} \right).$ These results are well-known in the literature and shown in \citet{rudolph2017robust} and \citet{rudolph2023efficiently}.
 Proof of this result as well as all the subsequent results are in Appendix~\ref{sec: proof_prelim}. Next, we focus on the estimation of TATE.

%% file: prelim/estimate.tex
\paragraph{Estimation.} 
For ease of illustration, we introduce our approach for a simple inverse probability-weighted (IPW) estimator for the TATE proposed in \citet{stuart2011}. A similar and analogous analysis can be carried out using any consistent estimator of TATE including an outcome model-based approach and a doubly robust approach combining outcome model and IPW approaches. Let $\yhat^{ipw}_i(t) = \frac{\phat}{1-\phat} \frac{S_i \mathbbm{1}(T_i = t) Y_i}{\lhatXi \ehat(\bX_i)}$. Then
\begin{align*}
    \attehat^{ipw} &= \widehat{\bar{Y}}^{ipw}(1) - \widehat{\bar{Y}}^{ipw}(0) \textrm{ where, } \widehat{\bar{Y}}^{ipw}(t) = \frac{1}{n_1} \sum_{i=1}^{n} \yhat^{ipw}_i(t).
\end{align*}
Let $\sigma^2(\bX,t) := \E[ (Y - \muX{t})^2 \mid \bX , T=t, S=1 ]$. Then, given assumptions \ref{a: positive} - \ref{a: external_val}, and unbiased estimator $\attehat^{ipw}$
\begin{equation}\label{eq: var_atte_sim}
    \sqrt{n_1}(\attehat^{ipw} - \atte) \overset{d}{\to} \mathcal{N}\left(0, \mathbb{V}^{ipw} \right)
\end{equation}
where 
\begin{eqnarray}\label{eq: var_atte_sim_var}
\mathbb{V}^{ipw} = \frac{\pi}{(1-\pi)^2}\E_{\bX|S=1} \left[ \left( \frac{\sigma^2(\bX, 1)}{\ell^2(\bX) e(\bX)} + \frac{\sigma^2(\bX, 0)}{\ell^2(\bX)(1-e(\bX))} \right)  + \frac{(1-\pi)^2}{\pi(\bX)} \left( \atte(\bX) - \atte \right)^2 \right].\end{eqnarray}
Further, the variance ${\mathbb{V}}^{ipw}$ is consistently estimated as follows:
$$\widehat{\mathbb{V}}^{ipw} = \frac{1}{n_1} \sum_{i=1}^n S_i \left(\yhat^{ipw}_i(1) - \yhat^{ipw}_i(0) - \attehat^{ipw}\right)^2$$ with $\widehat{\mathbb{V}}^{ipw} \overset{p}{\to} {\mathbb{V}}^{ipw}$ as $n\to\infty$. We show the derivation of asymptotic variance in Appendix~\ref{sec: proof_prelim}~(b).
\begin{remark}
An essential consideration in generalizing trial estimates lies in the precision of the effect estimates. As seen in Equation~\ref{eq: var_atte_sim_var}, the contribution to the variance is larger for smaller $\px$ i.e., the precision of estimates diminishes for the subspace exhibiting ``practical violations of the positivity'' \citep{petersen2012diagnosing}. 
A potential strategy for ensuring a precise estimate thus involves constraining the analysis to a subspace with larger $\px$. 
While this may enhance the precision of estimates, the drawback is the reduction in generalizability as well as the sample size $n_1$, as units with low $\px$ are pruned. This reduction in sample size can decrease the precision of the TATE estimate. 
\end{remark}

\begin{remark}
Now, let's shift our focus to the second term: $ (1-\pi)^2 \frac{\left( \atte(\bx) - \atte \right)^2}{\pi(\bx)} $. The numerator measures level of heterogeneity of the treatment effect i.e. how different is $\atte(\bx)$ from $\atte$, and the denominator $\px$, as before, measures how the representativeness of units with covariate $\bX=\bx$. If $\px$ is small, then it implies that there is limited overlap between the target and trial data in the region of covariate space with $\bX=\bx$. However, if $\atte(\bx)$ is not very different from $\atte$ then this limited overlap does not contribute high amount of variance or uncertainty in the estimation. This indicates that precise estimation of the TATE requires distributional alignment on covariates that are effect modifiers. This result is well known in the literature.\cite[e.g.,][]{rudolph2023efficiently,zeng2023efficient}
\end{remark}

%% file: goalpost.tex
\section{Shifting the Goalpost}\label{sec: framework}
Our study encompasses two interrelated objectives: (i) delineating subpopulation(s) that are present in the target but are underrepresented in the trial cohort, and (ii) enhancing the precision of estimation of the TATE. In pursuit of these goals, we propose a principled two-staged approach: (i) a design stage to characterize the population for which the trial evidence can be extended with high precision, and (ii) an analysis stage to estimate the target treatment effect for that population. 
As a direct consequence of the design stage, we also identify and characterize the underrepresented subpopulation, for which extrapolating causal effects may lead to inaccurate and/or imprecise estimates. 

We implement this by estimating  
a reweighted version of the TATE, which we 
refer to as the weighted TATE (WTATE):
\begin{equation}
    \atte^{w} = \frac{\p_w}{1-\p_w} \frac{\E[w(\bX) \atte(\bX)/\lX \mid S = 1]}{\E[w(\bX) \mid S = 1]},
\end{equation}
where, $\p_w = \E[w(\bX) S]$ and $\atte(\bX) = \E[Y(1) - Y(0) \mid \bX, S=0]$.
Here, a choice of \textit{binary} $w(\bx)$ corresponds to including or excluding the individuals with $\bX$ equal to $\bx$ in the study population. Analogously, it also corresponds to whether the individuals with covariate values $\bx$ are represented or underrepresented in the study population.

One of the goals of the design stage is to find a study population, defined by $w(\cdot)$, for which we can \textit{precisely} extend the trial evidence; the \textit{precision} is measured as the inverse of the variance of the estimate. The precision of an estimate is a function of: (i) the data generative mechanism, (ii) trial sample size, and (iii) estimator. While observed data along with the choice of $w(\cdot)$ defines our study population and the corresponding effective sample size, the choice of the estimator is an analytic choice. In what follows, we illustrate our proposed approach for the IPW estimator for the sake of simplicity and ease of communication. However, our approach could be similarly adapted for an outcome modeling-based approach, a doubly robust estimator, or for that matter, any other consistent estimator. The corresponding IPW estimator for WTATE is given as follows:
\begin{align*}
    \yhat^{ipw,w}_i(t) &= \frac{\phat_w}{1-\phat_w} \frac{S_i \mathbbm{1}(T_i = t) Y_i}{\lhatXi \ehat(\bX_i,t)}\\
    \attehat^{ipw,w} &= \widehat{\bar{Y}}^{ipw,w}(1) - \widehat{\bar{Y}}^{ipw,w}(0)\\& \textrm{ where } \widehat{\bar{Y}}^{ipw,w}(t) = \frac{1}{\sum_{i=1}^{n} w(\bX_i)S_i} \sum_{i=1}^{n} w(\bX_i) \yhat^{ipw,w}_i(t).
\end{align*}
$\attehat^{ipw,w}$ is a consistent and asymptotically normal estimator of $\atte^{w}$ such that:
$$
    \sqrt{n_1}(\attehat^{ipw,w} - \atte^{w})^2 \overset{d}{\to} \mathcal{N}\left(0, \mathbb{V}^{ipw,w}  \right)
\text{ where } {\small
\mathbb{V}^{ipw,w} := \frac{\p_w}{(1-\p_w)^2}
\frac{\E_{\bX\mid S=1}\left[ w^2(\bX) \alpha^2(\bX)\right]}{\E^2_{\bX\mid S=1}[w(\bX)]} }, $$
and $\alpha^2(\bX) =  \left( \frac{\sigma^2(\bX, 1)}{\ell^2(\bX) e(\bX)} + \frac{\sigma^2(\bX, 0)}{\ell^2(\bX)(1-e(\bX))} \right)  + \frac{(1-\pi)^2}{\pi(\bX)} \left( \atte(\bX) - \atte^w \right)^2.$
Here, the problem of characterizing the target population where we can precisely estimate the TATE reduces to finding optimal $w$ such that
\begin{equation*}
    w^{ipw,*} \in \arg \min_{w} \mathbb{V}^{ipw,w}.
\end{equation*}
In practice, we only have a finite sample $\setSn$ and $\mathbb{V}^{ipw,w}$ is consistently estimated by
\begin{equation*}
    \widehat{\mathbb{V}}^{ipw,w}_{\setSn} = \frac{n_1}{(\sum_i S_i w(\bX_i))^2} \sum_i w^2(\bX_i)S_i \left(\yhat^{ipw,w}_i(1) - \yhat^{ipw,w}_i(0) - \attehat^{ipw,w}\right)^2.
\end{equation*}
Thus, to characterize the study population for which we can precisely estimate treatment effects, we aim to find an optimal $w$ such that 
\begin{equation*}
    w^{ipw,*}_{\setSn} \in \arg \min_{w} \widehat{\mathbb{V}}^{ipw,w}_{\setSn}.
\end{equation*}
Not only do we want to learn an optimal function $w$ using the data $\setSn$ but also we want $w$ to be ``interpretable'' to ensure that we can decribe who is and is not included in the study population. Modeling $w$ as a function of summaries produced by $\ell$ and $\mu$ would not yield the interpretability needed. We model $w$ directly as a function of the covariates, $\bX$. 

%% file: estimation.tex
\section{Learning $w$ via Functional Optimization}\label{sec: method}
This section delineates our methodology for learning a (binary) function $w(\cdot)$ that minimizes the objective function of interest. Our overarching goals encompass identifying a study population conducive to precise estimation of target treatment effects and conveying the characteristics of that refined target population along with the underrepresented subgroups in the original target population but excluded from the refined target population. While the first goal frames a functional optimization problem, the second goal imposes interpretability constraints on the resulting potential $w$ functions. In this section, we present the optimization methodology in its generality for simplicity, reserving the discussion on optimizing in terms of TATE variance minimization for the end of the section.

Consider any arbitrary dataset $\setAn = \{\ba_1, \dots, \ba_n\}$, where $\ba_i = (\bc_i, \bv_i)$ comprises a $p$-dimensional real-valued covariate vector $\bc_i$ and a $k$-dimensional vector $\bv_i$, and an objective function $\mathcal{L}$ that operates on a dataset $\setAn$. Additionally, consider a binary function $w: \mathbb{R}^p \to \{0,1\}$, such that $\forall (\setAn, w)$, $\mathcal{L}(\setAn, w) \geq 0$. Our goal is to determine an optimal function $w$ that minimizes $\mathcal{L}$,
$ w^* \in \arg \min_{w} \mathcal{L}(\setAn, w) $
where $w^*$ is the optimal but unknown function, and our additional objective is to interpretably characterize $w^*$. 

To search for the optimal $w$, one option involves restricting the functional class of $w$, such as linear functions $w(\bc) = \bc \beta^T$, leading to a numerical optimization problem of finding optimal $\beta$. However, this approach might be limiting when the true $w^*$ does not belong to the restricted functional class. Another option is a fully nonparametric approach, letting $w(\bc) = \sum_{i=1}^{n} w_i \mathbbm{1}[\bC_i = \bc]$ — a weighted sum of sample indicators. This reduces the functional optimization problem to a numerical optimization task of finding an optimal $\{w_i\}_{i=1}^{n}$, an n-dimensional vector. However, as the dataset size $n$ increases, computational complexity escalates, making it less practical. Moreover, this nonparametric approach is prone to overfitting and lacks interpretability, making it challenging to communicate the characteristics of the study population.

Our proposed methodology models $w$ as a tree or a forest (an ensemble of trees), offering full nonparametric flexibility. Unlike the previous options, this approach involves no explicit numerical parameters to minimize. Further, an uninformed or brute-force search might not be effective here, as the space of potential trees can be significantly larger than even $\{0,1\}^n$. Instead, we simplify this problem to the iterative construction of interpretable trees in a way that aligns with the objective of minimizing our target objective function. In what follows, we first discuss the challenges in this problem setup that makes it different from the traditional tree-construction algorithms like CART \citep{breiman2017classification}, BART \citep{chipman2010bart}, or GODST \citep{lin2020generalized}. Then, we describe our tree-constructing algorithm that allows us to characterize an almost-optimal $w$.

Unlike traditional tree construction algorithms, we lack access to the outcomes from $w^*$; in other words, we don't have labels or outcomes that we aim to predict. Moreover, the conventional setup often computes the global loss on the full dataset $\setSn$ as the weighted average of losses for independent subsets, given by $\mathcal{L}(\setAn,w) = \alpha_A \mathcal{L}(\mathcal{D}_n^A,w) + \alpha_B \mathcal{L}(\mathcal{D}_n^B,w)$. Here, $\alpha_A \geq 0$ and $\alpha_B \geq 0$ are constants, $\mathcal{D}_n^A \cup \mathcal{D}_n^B = \setSn$, and $\mathcal{D}_n^A \cap \mathcal{D}_n^B = \varnothing$, allowing for independent optimization in different parts of the feature space. However, the objective function does not exhibit the same additivity property in our case. Therefore, during the tree construction process, we need to keep track of the global objective function as we split and partition the dataset. This distinctive characteristic of our setup requires a more intricate strategy, departing from the conventional divide-and-conquer or greedy approaches in tree optimization.

\subsection{Rashomon set of Optimal Trees (ROOT)}
Our approach adopts a randomization-based strategy where the choice of variables to split on is made randomly, deviating from the traditional greedy approach. This randomness is particularly relevant to the nature of our problem. The splitting routine in our approach takes as input the local partition of the dataset $\setAn^A = (\bC^A, \bV^A)$ based on prior splits (if any), a copy of the entire dataset $\setAn$, and a $p+1$ vector $\mathbbm{f}$. $j$-th entry in $\mathbbm{f} = \{ f_1, \dots, f_{p+1} \}$ for any $j \leq p$ corresponds to the probability of splitting on feature $B_j$, and the $p+1$-th entry corresponds to sampling a leaf and stopping the growth of the specific branch of the tree. To ensure the sparsity of the tree, we let $f_{p+1}$ be a relatively larger number compared to other features.

If a leaf is drawn, our approach sets $w(\bC) = c$ for all $\bx$ in that leaf. Here, $c$ is a random value from the range of $w$ with probability $\epsilon_{explore}$, and a greedy optimal value that reduces the objective function at the given level with probability $1 - \epsilon_{explore}$. This explore-exploit strategy helps avoid local minima \citep{iyer2023wide}. Instead of a leaf, if the $j$-th feature is drawn, we split at the midpoint of the empirical distribution of $B_j^A$. For the subsequent left and right partitions, we \textit{tentatively} set $w(\bc) = c^{left}$ for units in the left subtree and $w(\bc) = c^{right}$ for units in the right subtree, where $c^{left}$ and $c^{right}$ are greedy optimal values that reduce the objective function at the given level. If the global objective value is lower than the objective value beforehand, then we make the tentative assignments permanent. This process is followed recursively until we sample a leaf on each of the branches. The steps are delineated in more detail in Algorithm~\ref{alg:split}. 

Given the stochastic nature of our tree construction method, there is no assurance that a single tree is optimally minimizing the objective function. To address this challenge, we generate a set of $M$ trees, with a higher $M$ more desirable. For practical considerations, we construct approximately $M = 5000$ trees in each of our empirical studies, both with synthetic and real data. We then rank these trees based on their respective objective function values and select the top $m$ trees with the smallest objective values. We shape our final $w$ by aggregating predictions from each tree using a maximal voting approach. To ensure interpretability, we keep the value of $m$ relatively small, typically ranging between 5 and 15. We call this ensemble of selected trees the \textbf{R}ashomon set \textbf{O}f \textbf{O}ptimal \textbf{T}rees (ROOT). Although each tree in ROOT is nearly equally effective, they provide slightly different interpretations. Leveraging the sparsity of these trees and the manageable size of ROOT, we further create a concise (albeit non-optimal) single-tree \textit{explanation} for the $w$ characterized by ROOT. This single-tree explainer learns a sparse representation that mimics the prediction function learned by ROOT -- we also refer to this as the \textit{characteristic tree}. We delineate the algorithm in Appendix~\ref{sec: root_alg}.

We operationalize the proposed ROOT approach for our setup where we are interested in identifying which units to include in our analysis to obtain the most precise estimate of TATE. For our problem, $\setAn = \setSn$, $\bC_i = \bX_i$, $\bV_i = (S_i, T_i, Y_i)$ and $\mathcal{L}(\setSn,w)$ is the estimate of the finite sample variance $\widehat{\mathbb{V}}^{ipw,w}$ of $\sqrt{n_1}(\attehat^w - \atte^w)$ where $$\widehat{\mathbb{V}}^{ipw,w} = \frac{n_1}{\left(\sum_{i=1}^{n} S_i w(\bX_i)\right)^2} \sum_{t\in\{0,1\}} \sum_{i=1}^{n} S_i \left(w(\bX_i)\right)^2  \left( \widehat{Y}_i^{ipw}(t) - \widehat{\bar{Y}}(t) \right)^2,$$ and $n_1 = \sum_{i=1}^{n} S_i.$

%% file: moud_trial.tex
o\section{Empirical Results}
\subsection{Summary of Synthetic Data Analyses}
We conducted synthetic data experiments to evaluate the effectiveness of our proposed approach, ROOT, in generating interpretable representations while accurately characterizing underrepresented groups. These experiments were designed to assess ROOT’s performance in high-dimensional and non-linear settings, comparing it to alternative methods. We employed four distinct data generative procedures, including one that closely mimics real-world data on medications for opioid use disorder.  Detailed descriptions of the data generative procedures and the subsequent analyses are provided in Appendix~\ref{sec: synth_exp}. As shown in Table~\ref{tab: synth_exp}, refining the study population using ROOT led to significant improvements in precision. Additionally, Figure~\ref{fig: synth_tree} illustrates that ROOT produces interpretable results in the form of sparse or shallow decision trees. However, while ROOT performs optimally across most data generating processes (DGPs), it struggles to maintain sparsity when the effect modifiers themselves are high-dimensional. This limitation highlights a potential area for future refinement.

\subsection{Insufficiently Represented Groups in MOUD Trials}\label{sec: moud}
Now, we apply our methodology to the medication for opioid use (MOUD) data where we are interested in transporting the treatment effect from the Starting Treatment With Agonist Replacement Therapies (START) trial to the population of individuals in the US seeking treatment for opioid use disorder, using the Treatment Episode Dataset-Admissions 2015-2017. We are interested in (i) the characteristics of the subpopulation for which we can precisely estimate the TATE using the trial evidence, (ii) the TATE estimate for this subpopulation, and (iii) the characteristics identifying those who are treated in TEDS-A but underrepresented in the trial cohort.
\subsubsection{Data Description}
We first begin the discussion by describing the two datasets used:  the trial and population data.

\paragraph{Trial Data.} 
The Starting Treatment With Agonist Replacement Therapies (START) trial, initiated in April 2006, was a multi-center study designed to compare the effects of buprenorphine and methadone \citep{saxon2013buprenorphine, hser2014treatment}. It involved 1271 participants who were randomly assigned to either the buprenorphine or methadone arms in a 2:1 ratio. We chose this trial as it is one of the largest MOUD trials conducted as part of the NIDA Clinical Trials Network, which are trials that are conducted in real-world treatment settings, designed to improve generalizability \citep{tai2011national}. The trial's findings indicated that methadone was associated with better retention in treating opioid use disorder compared to buprenorphine \citep{hser2014treatment}. However, it must be noted that the trial compares methadone to an atypical buprenorphine administration, deviating from real-world practices. Individuals had to return to the study site daily for distribution of buprenorphine.  In usual care settings, buprenorphine is distributed on a 2-6 week prescription interval .

In our analysis, the outcome is relapse by 24 weeks following the treatment assignment. \textit{Relapse} is delineated as the consistent use of non-study opioids at least once per week for four consecutive weeks or daily usage of non-study opioids for seven continuous days. Data on opioid use is collected through urine drug screens and self-reported during follow-up visits. The assessment of relapse initiates 20 days post-randomization to account for positive urine screens that may be attributed to opioid use during medically managed withdrawal or residual drug presence during stabilized treatment.

\paragraph{Target Sample.} We employed the Treatment Episode Dataset - Admission (TEDS-A) dataset from 2015-2017 as a representative cross-section of the target demographic. This dataset includes individuals who sought assistance from publicly funded substance use treatment programs, spanning inpatient, outpatient, and residential facilities under the management of the Substance Abuse and Mental Health Services Administration. The TEDS-A dataset utilized in this study comprised information from 48 states (excluding Oregon and Georgia) and the District of Columbia. Notably, data for South Carolina was unavailable for the year 2015. Our analysis focuses on a cohort of 740,015 individuals who fulfilled two specific criteria: (i) the inclusion of Medication-Assisted Treatment (MOUD) in their treatment plan and (ii) a DSM-IV-TR diagnosis of opioid dependence or opioid abuse, or individuals entering treatment with documented heroin use, non-prescription methadone use, or other/synthetic opioid use.

\paragraph{Data Summary.} For our analysis, we consider methadone as $T=1$ and buprenorphine as $T=0$. For the outcome, $Y=1$ indicates relapse by 24 weeks after treatment assignment. We only consider the pre-treatment covariates observed in both the trial and target data. Our set of pre-treatment covariates includes the participants' age, race and biological sex along with substance use history (amphetamine, benzodiazepines, cannabis and IV drug use) assessed at the initiation of MOUD treatment. See Table~\ref{tab: data_moud} for a listing of all of the covariates included.

\subsubsection{Analysis}
\paragraph{Trial ATE.} We first estimate the average treatment effect in the trial sample. Individuals assigned to methadone exhibit a $10.04$ percentage point reduction in the likelihood of relapse by 24 weeks compared to those on buprenorphine i.e. the experimental ATE is $-10.04$\footnote{This particular comparative treatment effect compares methadone to an atypical administration of buprenorphine in which buprenorphine is administered during daily or near-daily clinic visits to match the administration of methadone, but which is markedly different from administration in the real world in which buprenorphine is prescribed every 2-6 weeks. Therefore, this particular comparison involves a much more burdensome administration of buprenorphine.}, with a standard error of $3.66$.

\paragraph{Estimating TATE.} Next, we employ the IPW estimator to extend the inferences from the START trial to the population of opioid users represented by the TEDS-A dataset.  This relies on assumptions~\ref{a: positive}-\ref{a: sutva} We discuss the validity of these assumptions in Appendix~\ref{sec: discuss_assume}. The TATE estimate using IPW indicates a modest advantage for methadone, with a $-9.72$ percentage points reduction in relapse likelihood, albeit with a wide standard error of $4.680$. Note that the estimated TATE has a relatively larger standard error than the trial sample average treatment effect.

\paragraph{Shifting the Goalpost.} Now, we focus on refining the analysis by employing a two-stage approach: (i) the design stage identifies the underrepresented subpopulations and subsequently excludes them from the study population, and (ii) the analysis stage estimates the TATE for this refined study population. For the design stage: we use four different approaches including our proposed one using ROOT, two traditional approaches using selection scores ($\ell(\bx)$) and using ``indicator'' $w(\bx)$, and compare the results using each of them.
For the analysis stage: we use the IPW estimator to estimate TATE using the refined target population. We compare the results via ROOT with the results using the selection score. 
\begin{figure}
    \centering
    \includegraphics[width=\textwidth]{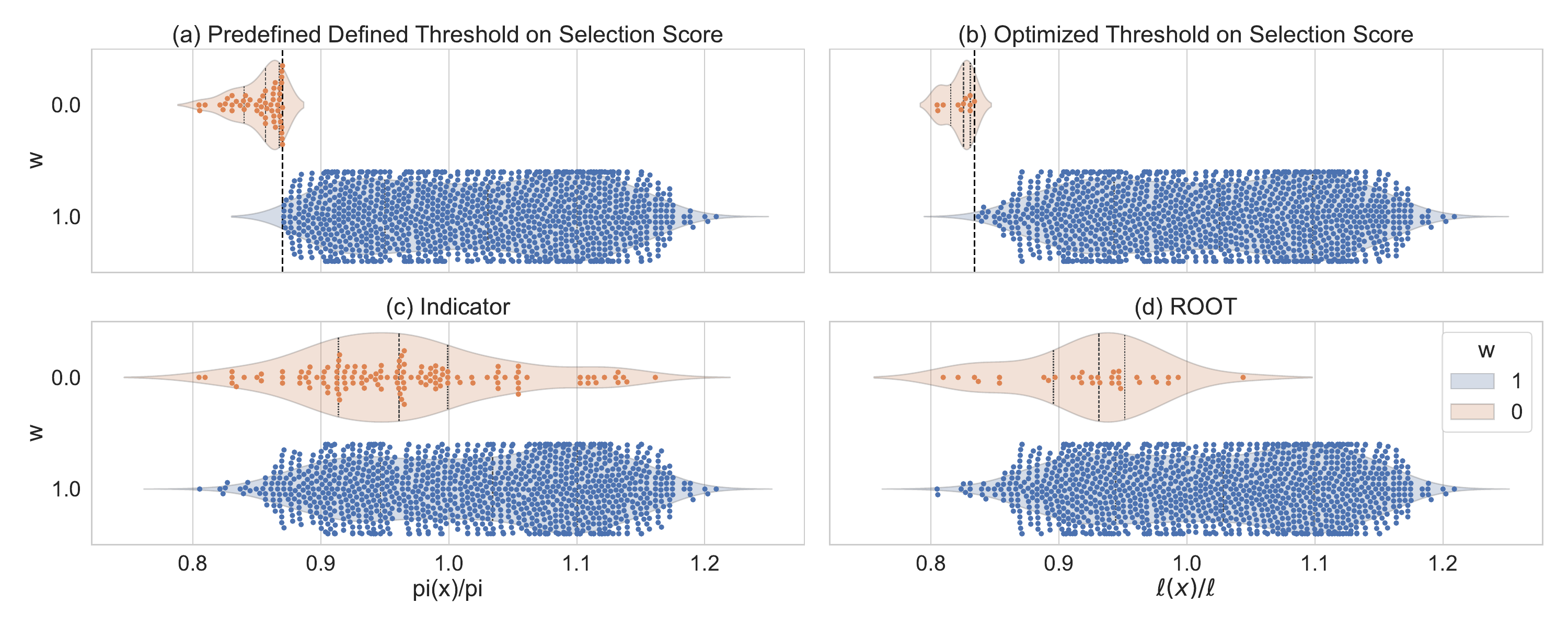}
    \caption{Swarm-plot and violin-plot showing the distribution of the normalized selection odds $\ell(\bx)/\ell$ for the units in the trial sample ($S=1$) characterized as $w(\bx)=0$ (i.e. ``underrepresented'') and $w(\bx)=1$ using (a) selection odds with predefined threshold of $0.87$ on $\ell(\bx)/\ell$, (b) selection odds with the optimal threshold for $\ell(\bx)/\ell$  and (c) using ROOT.}
    \label{fig: underrep_root}
\end{figure}

The traditional selection score-based approaches are akin to an approach proposed in \citet{crump2009dealing} and \citet{li2019addressing} to identify underrepresented populations using $\ell(\bx)$, and in particular the normalized selection odds, i.e., $\ell(\bx)/\ell$. In this approach, the problem reduces to finding a threshold $\ell_{\spadesuit}$ such that units with $\ell(\bx)/\ell$ less than $\ell_{\spadesuit}$ are underrepresented. We consider two versions of such a selection score-based ``design-stage'': the first version uses a pre-defined $\ell_{\spadesuit} = 0.87$ (we select this using the natural breakpoint in the distribution of normalized selection odds), and the second version learns the optimal $\ell_{\spadesuit}$ through an  ``indicator'' $w(\bx)$ approach, which chooses $w(\bx) = \sum_i w_i \mathbf{1}[\bx_i = \bx]$ and optimizes for $\{w_i\}$. Lastly, we use our proposed approach, ROOT, to construct $w(\bx)$ using trees.

\begin{figure}
    \centering \includegraphics[width=0.75\textwidth]{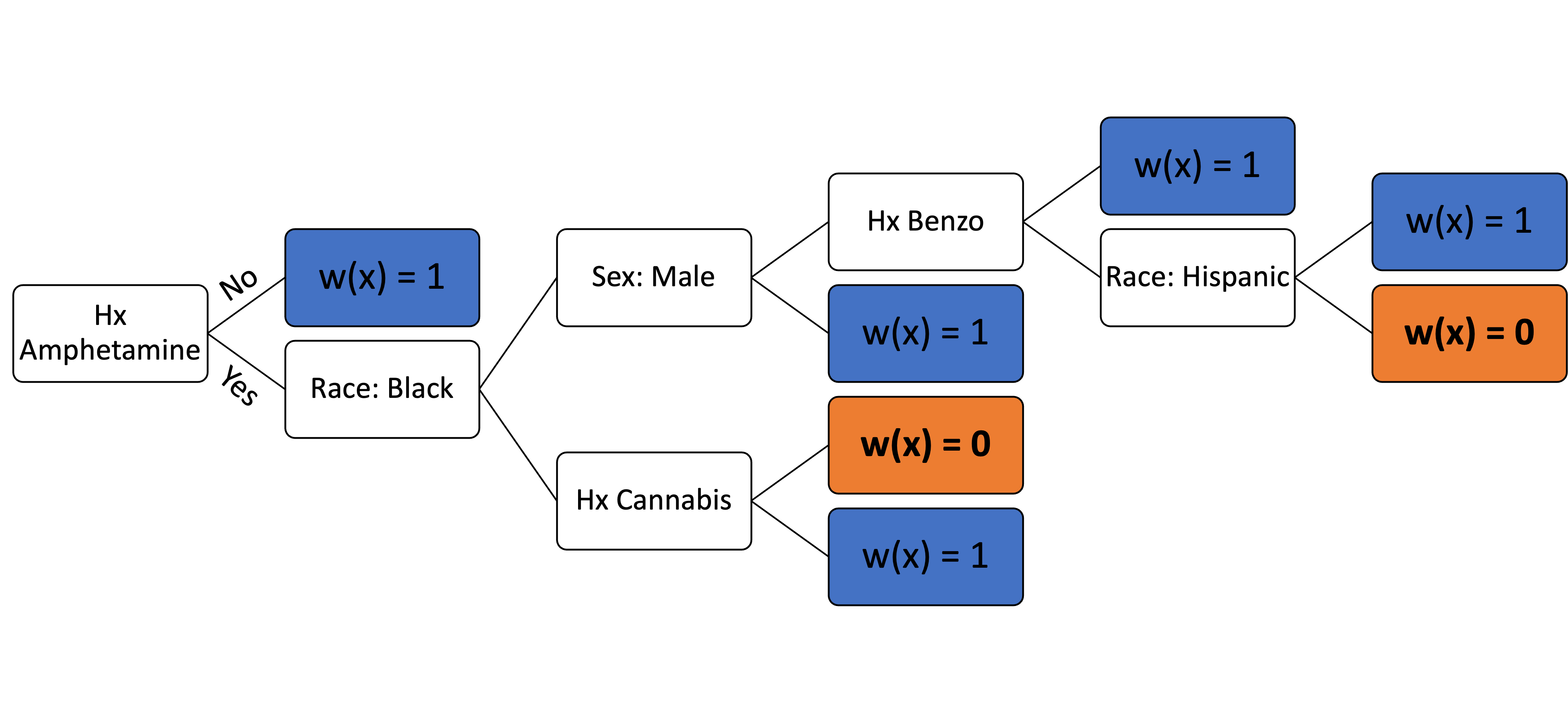}
    \caption{Decision Tree characterizing the underrepresented population. The orange nodes and leaves indicate underrepresented subgroups that can be excluded from the TATE analysis for better precision.}
    \label{fig: underrep_tedsa}
\end{figure}

Table~\ref{tab: summary_est} shows the WTATE point estimate and standard errors for each of the four design stage strategies along with the original TATE estimates and sample ATE. Figure~6 shows the distribution, along $\ell(\bx)/\ell$, of the refined target population and the subpopulations characterized as underrepresented for each of the four approaches. The ``indicator'' $w(\bx)$ yields the WTATE with the smallest standard error followed by the one using ROOT. However, ``indicator'' $w(\bx)$ results in pruning many more units from the target population compared to ROOT, thus limiting the population for inference. Further, ``indicator'' $w(\bx)$ also characterizes a significant subpopulation with $\ell(\bx)/\ell$ greater than 1 as underrepresented -- this is potentially a sign of overfitting. Overall, ROOT produces smaller standard errors compared to both the selection score-based approaches (third column of Table~\ref{tab: summary_est}) while also ensuring that the refined target population is as large as possible (last column of Table~\ref{tab: summary_est}), avoiding overfitting (last column of Table~\ref{tab: summary_est}) and ensuring interpretability (Figure~\ref{fig: underrep_tedsa}).

\begin{table}[]
\caption{Table of estimands corresponding point estimate, standard error, and relative size of the refined target population measured via $\E[w(\bX)]$ for the four choices of design stage strategies.}
\label{tab: summary_est}
{\begin{tabular}{lcclc}\\
\hline
    \textbf{Estimand} & \textbf{Point Est} & \textbf{Std Err} & \textbf{Remarks} & $\E[w(\bX)]$\\
\hline
    Trial ATE & -10.040 & 3.666 & on START trial sample & - \\
    TATE & -9.724 & 4.680 & using full TEDS-A data & 1.000 \\
    WTATE (selection score) & -8.641 & 4.750 & predefined threshold & 0.884\\
    WTATE (selection score) & -10.171 & 4.676 & optimized threshold & 0.938\\
    WTATE (Indicator) & -27.744 & 4.612 & - & 0.888 \\
    WTATE (ROOT) & -12.182 & 4.666 & - & 0.993\\\hline
\end{tabular}}
\end{table}

\paragraph{Characterizing Underrepresented Populations.}
One key aspect of ROOT's design stage is its interpretability along with flexibility, which allows for communication of the characteristics of the underrepresented population that is excluded from the refined study population. Figure~\ref{fig: underrep_tedsa} presents a decision tree that characterizes these subgroups. We identify that the following two subgroups are underrepresented in the START trial: (i) Black people with a history of amphetamine use but no history of cannabis use, and (ii) Hispanic women with a history of amphetamine and benzodiazepine use. 

\begin{figure}
    \centering
    \begin{tabular}{cc}
    \includegraphics[width=0.4\textwidth]{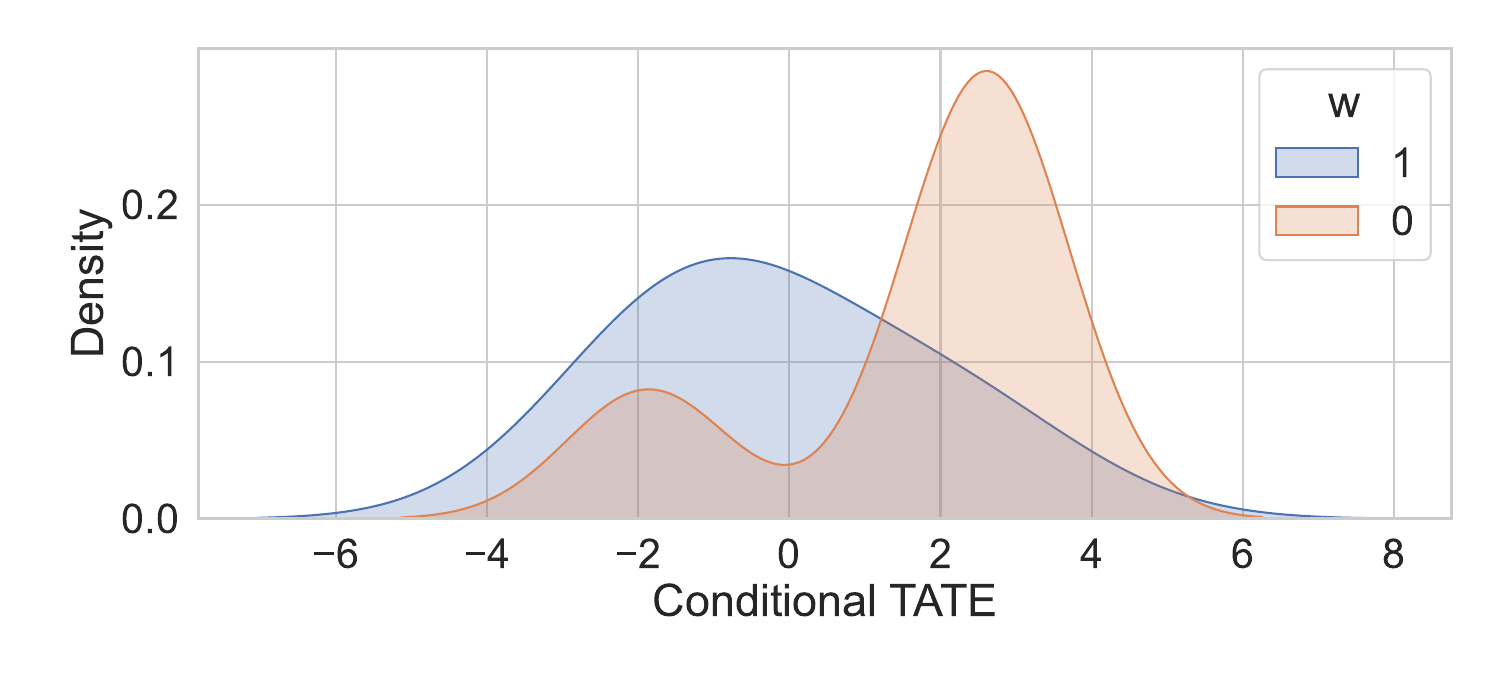}  & \includegraphics[width=0.56\textwidth]{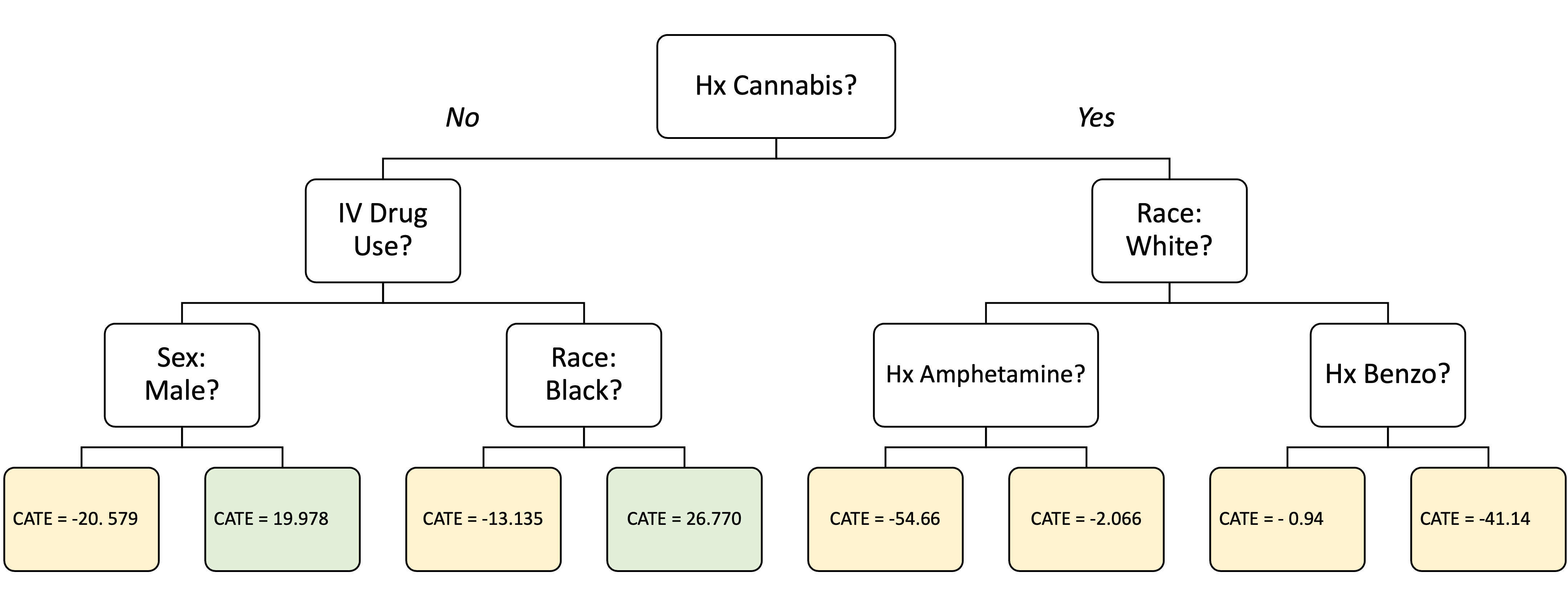} \\
         (a) & (b) 
    \end{tabular}
    \caption{(a) Distribution of conditional average treatment effects for units with $w(\bX_i)=1$ (well represented) and $w(\bX_i)=0$ (characterized underrepresented) via ROOT. (b) Tree explaining the distribution of conditional average treatment effects across various subgroups. Here, the yellow leaves indicate methadone is more effective compared to buprenorphine while the green leaves indicate buprenorphine is more effective than methadone in that subpopulation.}
    \label{fig:moud_cate}
\end{figure}

One key difference between these insufficiently represented groups and the remaining groups is the distribution of conditional average treatment effects (as shown in Figure~\ref{fig:moud_cate}(a). Figure~\ref{fig:moud_cate}(b) performs a deeper dive into understanding the distribution of conditional average treatment effects as a function of covariates. Our results indicate that while for the majority of the population methadone (T=1) appears to be the superior treatment, for the following two subgroups buprenorphine (T=0) is the superior treatment: individuals who do not have a history of cannabis use and are either (i) racially Black or (ii) biological males. Of these two groups, the average normalized selection odds, $\ell(\bx) / \ell$, and corresponding CATE estimates for racially Black individuals is $0.909$ and $+26.770$, and for the biological males is $1.001$ and $+19.978$, respectively. The heterogeneity of CATE along with the selection odds less than one implies a potential lack representation on effect modifiers. These analysis results are in agreement with the ROOT-based characteristic tree which identifies Black individuals without cannabis use history as insufficiently represented in the START trial.

%% file: conclude.tex
\section{Discussion \& Conclusion}\label{sec: conclude}
Our paper addresses the common issue of insufficiently represented (also referred here as underrepresented) populations in clinical trials when generalizing treatment effects to a target population. The presence of insufficiently represented subgroups can significantly undermine the accuracy and precision of estimated target treatment effects, thereby impacting subsequent decision-making processes. In response to this challenge, we propose a principled approach to characterize the underrepresented groups by optimizing for precision systematically. This facilitates a clear understanding of subpopulations for which precise estimation of treatment effect is difficult. Such an exercise of interpretable characterization is of value for but not limited to (i) understanding contributions to research-to-practice gaps, such as in medication for opioid use disorder studies, and (ii) designing future studies to understand effects for the underrepresented groups.

Our methodology introduces a weighted estimand, where these weights, determined by pretreatment covariates, guide the inclusion or exclusion of units in the refined target population. We restrict these weights to binary values, indicating the selection or exclusion of units from the analysis. Leveraging available trial data, we introduce a novel functional optimization approach, the Rashomon set of optimal trees (ROOT), to learn these binary weights and minimize the variance of the estimated target average treatment effect. ROOT stands out for its interpretability and flexibility.

\paragraph{On Refined Estimand.} Our procedure redefines the estimand to focus on a causal quantity within a well-defined subgroup that has sufficient support in the data. The primary goal is to identify and characterize underrepresented regions of the covariate space in the trial cohort. Once these regions are identified, they are excluded from our analysis, which means dropping units in these regions from both the experimental and target samples. Consequently, the refined population is likely not identical to the original trial cohort. This approach involves a fundamental tradeoff: while our generalized estimates for the refined estimand, WTATE, have greater credibility, they are less general than the original estimand, TATE. Note that while it is crucial to ensure balance on effect moderators when generalizing or transporting inferences, extrapolating in the direction of features that are not effect modifiers is feasible without any loss of precision.

\paragraph{On Interpretability.} 
Sparse decision trees are learned across the literature for their ability to produce interpretable results. However, there are possibly other decision trees that are nearly as optimal as the learned tree and have an alternate interpretation \citep{xin2022exploring, semenova2022existence}. Thus, analogous to the traditional interpretable ML literature \citep{xin2022exploring, ruggieri2019complete}, we use the Rashomon set of the optimal trees as the interpretations are more comprehensive and stable. Here, the Rashomon set is constrained to contain the $k$ best trees with the smallest objective value out of $M$ generated trees (e.g., in our case study we choose $k=5$). The Rashomon set itself is considered ``interpretable'' because it is small and each element is individually interpretable. However, some  researchers  may desire a simpler characterization. Thus, we also summarize (aggregate) the representations produced by the trees in the Rashomon set via a characteristic tree, providing a result that is both interpretable and simple, though containing less expressive power than the full Rashomon set of optimal trees. Our software package outputs both this characteristic tree and the full Rashomon set with the respective objective values for each tree. 


\paragraph{Takeaways from Empirical Study.} Our synthetic data experiments demonstrate that in scenarios where the number of moderators is relatively small, our approach enhances the precision of estimates compared to alternative methods while maintaining interpretability. We also explore the limits of our proposed approach via a high-dimensional scenario with a large number of moderators where it is challenging to construct a sparse representation of the underrepresented population. Lastly, we also compare the performance of ROOT and other baseline approaches on a synthetic DGP that mimics the distribution of our real-world MOUD case study data. Here, ROOT outperforms other methods as well as yields an interpretable representation of the underrepresented population that is in congruence with the DGP. Furthermore, we apply our method to extend inferences from the START trial, generalizing the comparative treatment effect of methadone vs. buprenorphine on OUD relapse within 24 weeks after randomized assignment. Our analysis identifies underrepresented subgroups in the trial using a shallow, easily interpretable tree. Post-pruning, our refined analysis significantly improves the precision of the generalized effect estimate.

\paragraph{On Data-Adaptive Estimand.} While our approach significantly enhances the quality of analyses and subsequent decision-making, it also highlights open questions. For example, similar to \cite{crump2009dealing, li2019addressing}, we use the data to define the estimand, which has implications for inference. We conceptualize our approach as a two-step procedure, encompassing design and analysis. In the design step, we leverage ROOT alongside trial data to refine the target population. In the subsequent analysis step, we consider this redefined estimand fixed and proceed to estimate the target average treatment effect. An open question pertains to whether accounting for uncertainty in estimating the weights $w$ is necessary and how it impacts TATE estimation and inference.

\paragraph{On Positivity Violation} 
For consistent estimation of the target average treatment effect, representativeness is required only on effect moderators that differ in distribution between the trial and target, rather than on every feature. However, in many real-world scenarios, we do not know which features act as effect moderators. Our procedure leverages observed data to implicitly identify these effect moderators. In this context, our method assumes positivity and focuses on identifying ``underrepresented'' populations -- those that are present but not sufficiently so -- rather than completely ``unrepresented'' populations. To illustrate this with an example, consider a binary feature encoding ``pregnancy'': if the trial includes only non-pregnant individuals and excludes pregnant individuals, we cannot ascertain if pregnancy status is an effect modifier using the observed data alone. Consequently, we cannot determine the necessity of balancing pregnancy status or if pregnant individuals are underrepresented. Thus, we recommend preprocessing the data to a priori prune regions with positivity violations (i.e., $P(S=1 \mid \bX) = 0$) based on domain knowledge or trial admission/exclusion criteria.

\paragraph{On Existence of Unobserved Confounder.} 
In addition to positivity, generalizability or transportability also depends on the external validity of RCT (A3) --- this assumption is also referred to as S-admissibility or sample exchangeability. There is an extensive literature that focuses on the partial identification of transported/generalized treatment effects and sensitivity analysis of the estimation step to violations of the external validity assumption \citep{dahabreh2022global,colnet2021generalizing, nguyen2017sensitivity, huang2024sensitivity, zeng2023efficient, nie2021covariate}. However, to the best of our knowledge, no prior work has discussed the sensitivity of a data-adaptive estimand, like the one we propose here, to a violation of the external validity assumption. This is a key gap, as mismeasurement or the absence of an important effect moderators can change the eventual estimand of interest. Future work could focus on developing sensitivity analysis tools for data-adaptive estimands such as that proposed here.

\paragraph{On Optimality Guarantee of ROOT.} Furthermore, while ROOT is employed for estimating $w$, it is an algorithm that uses randomization, and its general optimality is not guaranteed. In Appendix~\ref{sec: appendix_theory}, we show that ROOT's proposed solution achieves optimality, given a fixed dataset, as the number of grown trees approaches infinity, representing an exhaustive exploration of the entire tree space. However, for a finite set of trees, the optimality gap remains unknown. Future work will concentrate on exploring the computational and algorithmic properties of ROOT, aiming to quantify this optimality gap and quantify sampling uncertainty. While ROOT is tailored for characterizing target populations and minimizing variance, its versatility extends beyond the presented application. Our ongoing work will focus on extending ROOT to a broader setting and adapting it for non-binary $w$ functions. Further, we will explore the convergence, stabilization and finite-sample optimality gap of ROOT-based optimization. 

%% file: appendix_proof_prelim.tex
\section{Proof of Results in Section~\ref{sec: prelim}} \label{sec: proof_prelim}
\paragraph{(a) Identification of ATTE}
Recall, the ATTE is defined as $\atte = \E( \tau_i | S_i = 0)$. If we substitute the definition of $\tau_i$ in the above definition then $\atte = \E( Y_i(1) - Y_i(0) | S_i = 0) = \E( Y_i(1) | S_i = 0) - \E( Y_i(0) | S_i = 0)$. Given the decomposition, identification of $\E( Y_i(t) | S_i = 0)$ for $t\in\{0,1\}$ guarantees identification of ATTE. Next, we show the identification of $\E( Y_i(t) | S_i = 0)$. For convenience, we drop the index $i$ from the following discussion, given that units are iid sampled from the population:
\begin{align*}
    \E(Y(t) | S=0) &= \E_{\bX|S=0}(\E(Y(t) | \bX, S=0))\\
    &\overset{\ref{a: external_val}}{=} \E_{\bX|S=0}(\E(Y(t) | \bX, S=1))\\
    &\overset{\ref{a: internal_val}}{=} \E_{\bX|S=0}(\E(Y(t) | \bX, T=t, S=1))\\
    &\overset{}{=} \E_{\bX|S=0}(\E(Y | \bX, T=t, S=1))\\
    &\overset{}{=} \E_{\bX|S=0}(\muXt)
\end{align*}
\begin{equation}
    \implies \atte = \E_{\bX|S=0}(\muX{1} - \muX{0}).
\end{equation}
For ATTE, an alternate and equivalent identification strategy is:
\begin{align*}
   \E_{\bX|S=0}(\muXt) &= \int_\bx \muxt p(\bx|S=0) d\bx\\
   &= \int_\bx \muxt \frac{p(\bx|S=0)}{p(\bx|S=1)} p(\bx|S=1) d\bx\\
   &= \int_\bx \muxt \frac{p(S=0 \mid \bx) p(\bx) P(S=1)}{P(S=1 \mid \bx) p(\bx) P(S=0)} p(\bx|S=1) d\bx\\
   &= \frac{\p}{1-\p}\int_\bx \muxt \frac{1-\px}{\px} p(\bx|S=1) d\bx\\
   &= \frac{\p}{1-\p}\int_\bx \frac{\muxt}{\lx}  p(\bx|S=1) d\bx\\
   &= \frac{\p}{1-\p} \E_{\bX \mid S=1} \left( \frac{\muXt}{\lX} \right) \\
\end{align*}
\begin{equation}
   \implies \atte = \frac{\p}{1-\p} \E_{\bX \mid S=1} \left( \frac{\muX{1}-\muX{0}}{\lX} \right)
\end{equation}


\paragraph{(b) Variance of TATE IPW Estimate $\attehat$}
\[
    \attehat = \frac{1}{n} \sum_{i=1}^n \psi_i
\text{ where } \psi_i = \left( \frac{S_i T_i Y_i}{e(X_i) \pi(X_i)} - \frac{S_i (1-T_i) Y_i}{(1-e(X_i)) \pi(X_i)} \right) \frac{(1-\pi(X_i))}{(1-\pi)}. \]
Now, consider
\[ B = \attehat - \atte = \frac{1}{n} \sum_{i=1}^n (\psi_i - \atte ) = \frac{1}{n} \sum_{i=1}^n (\psi_i - \atte(X_i) + \atte(X_i) - \atte ) \]
where \(\atte(X_i) = \frac{1-\pi(X_i)}{1-\pi} 
\left( \mu(X_i,1) - \mu(X_i,0) \right)
\)
.
Let \(\sigma^2(x, t)\) be the variance of \(Y\) conditional on \(X = x\), \(T = t\) and \(S = 1\).

\[ Var(\attehat - \atte) = \E[B^2] = \frac{1}{n} Var(\psi - \atte(X) + \atte(X) - \atte ) \]
\[ Var(\psi - \atte(X) + \atte(X) - \atte ) = \E_X[ \E[(\psi - \atte(X) + \atte(X) - \atte )^2 \mid X] ] \]
\[ = \E_X[ \E[(\psi - \atte(X))^2 \mid X] ] + \E_X[ \E[(\atte(X) - \atte )^2 \mid X] ]\]

Now, first let's consider \(\E_X[ \E[(\psi - \atte(X))^2 \mid X] ] \):
\[ = \E_X\left[ \E\left[ \left(\frac{1-\pi(x)}{1-\pi}\right)^2\left(  
\frac{STY}{\pi(x)e(X)} -\mu(X,1)  - \frac{S(1-T)Y}{\pi(x)(1-e(X))} + \mu(X,0) \right)^2 \mid X\right] \right]\]
\[ = \E_X \left[ \left( \frac{\sigma^2(X, 1)}{e(X) \pi(X)} + \frac{\sigma^2(X, 0)}{(1-e(X)) \pi(X)} \right) \left( \frac{(1-\pi(X))}{(1-\pi)} \right)^2 \right] \]
\[ = \frac{\pi}{(1-\pi)^2}\E_{X|S=1} \left[ \left( \frac{\sigma^2(X, 1)}{e(X)} + \frac{\sigma^2(X, 0)}{(1-e(X))} \right) \left( \frac{(1-\pi(X))}{\pi(X)} \right)^2 \right] \]

Focusing on \( \E_X[ \E[(\atte(X) - \atte )^2 \mid X] ] \):
\[ \E_X[ \E[(\atte(X) - \atte )^2 \mid X] ] =  \E_X\left[ \left( \frac{1-\pi(X)}{1-\pi}(\mu(X,1) - \mu(X,0)) - \atte \right)^2 \right]  \]

\[ =  \E_{X|S=1}\left[ \frac{\pi}{\pi(X)} \left( \atte(X) - \atte \right)^2 \right] \]

Thus,
\begin{eqnarray*}
Var(\attehat - \atte) &=& \frac{\pi}{(1-\pi)^2}\E_{X|S=1} \left[ \left( \frac{\sigma^2(X, 1)}{e(X)} + \frac{\sigma^2(X, 0)}{(1-e(X))} \right) \left( \frac{(1-\pi(X))}{\pi(X)} \right)^2 \right] \\&& + \E_{X|S=1}\left[ \frac{\pi}{\pi(X)} \left( \atte(X) - \atte \right)^2 \right]
\end{eqnarray*}

%% file: appendix_alg_root.tex
\newpage
\section{ROOT Algorithm}\label{sec: root_alg}
\input{algorithm}

%% file: algorithm.tex
\begin{algorithm*}[!h]
{\small
\caption{Splitting Algorithm for ROOT}\label{alg:split}
\begin{algorithmic}[1]
\Function{split}{$\mathbbm{f}$, $\bX^A$, $\mathcal{D}_n$, parent\_loss, w, depth}
    \State \Comment{Sample the feature $f_j$ to split on based on the current depth}
    \State $f_j \gets \text{choose}(\mathbbm{f}, depth)$

    \If{$f_j$ equals ``leaf''}
        \State $w'(\bX; c) \gets \mathbbm{1}[\bX \in \bX^A] c + \mathbbm{1}[\bX \notin \bX^A] w(\bX)$
        \State \Comment{Decide whether to exploit (choose best loss) or explore (random choice)}
        \State $c_{\text{exploit}} \gets \text{arg min}_c (\{ \text{loss}(w',\mathcal{D}_n) : \text{for } c \text{ in } \{0,1\} \})$
        \State $c_{\text{explore}} \sim \text{bernoulli}(1, 0.5)$
        \State $e \sim \text{bernoulli}(1, \epsilon_{explore})$

        \State \Comment{Combine exploration and exploitation to decide the final weight}
        \State $w(\bX) \gets (e \times w'(\bX; c_{\text{explore}})) + ((1 - e) \times w'(\bX; c_{\text{exploit}}))$

        \State \Return\\
            \hspace{25mm} \begin{tabular}{rlrlrl}
                 \{ 
                ``node'':& $f_j$; &
                ``objective'':& loss$(w,\setSn)$; &
                 ``depth'':& $depth$
            \}
            \end{tabular} 
    \Else
        \State $u_j \gets \text{midpoint}(X^A_j)$

        \State \Comment{Split the dataset into left and right based on the midpoint}
        \State $\bX^A_{left} \gets \{\bX^A_{i} \text{ where } X_{i,j} \leq u_j\}$
        \State $\bX^A_{right} \gets \{\bX^A_{i} \text{ where } X_{i,j} > u_j\}$

        \State \Comment{Calculate losses for left and right branches}
        \State $w'(\bX; c_{left},c_{right}) \gets \mathbbm{1}[\bX \in \bX^A_{\text{left}}]c_{left} + \mathbbm{1}[\bX \in \bX^A_{\text{right}}]c_{right} + \mathbbm{1}[\bX \notin \bX^A]w(\bX)$
        
        \State $l' \gets \{ \text{loss}(w', \setSn) : (c_{left},c_{right}) \in \{0,1\}^2 \}$

        \State \Comment{Calculate new loss after the split}
        \State new\_loss $\gets \min_{(c_{left},c_{right})} l'$

        \State \Comment{Check if the new split reduces loss compared to the parent node}
        \If{new\_loss $\leq$ \text{parent\_loss}}

            \State \Comment{Update $w$ function}
            \State $w \gets \arg \min_{w'} l'$

            \State \Comment{Recursively split the left and right branches}
            \State \Return \\
            \hspace{25mm} \begin{tabular}{rl}
                 \{ 
                ``node'':& $f_j$, 
                ``split'': $u_j$, \\
                ``left\_tree'':& \textsc{split}($\mathbbm{f}$, $\bX^A_{left}$, $\mathcal{D}_n$, loss$(w,\setSn)$, $w$, $depth+1$), \\
                ``right\_tree'':& \textsc{split}($\mathbbm{f}$, $\bX^A_{right}$, $\mathcal{D}_n$, loss$(w,\setSn)$, $w$, $depth+1$), \\
                ``objective'':& loss$(w,\setSn)$, \\
                 ``depth'':& $depth$
            \}
            \end{tabular}
        \Else
            \State $\mathbbm{f}' \gets \text{reduce\_probability}(f_j, \mathbbm{f})$
            \State \Return \textsc{split}($\mathbbm{f}'$, $\bX^A$, $\mathcal{D}_n$, parent\_loss, w, $depth$)
        \EndIf
    \EndIf
\EndFunction
\end{algorithmic}
}
\end{algorithm*}

%% file: appendix_theory.tex
\section{Consistency of Tree-based Optimization}\label{sec: appendix_theory}
In this section, we discuss the convergence of random tree sampling-based optimization such as ROOT. ROOT samples random trees that favor splits on features correlated with the outcome i.e. the probability of splitting on a feature is equal to the correlation between the feature and the outcome. However, here, we show proof for a pure random tree with $Q$ leaves (PRTQ) sampling approach that splits uniformly over each feature. This is done for mathematical simplicity, and the PRTQ sampling approach is a special case of ROOT. Furthermore, trees sampled via a pure random tree sampling approach are independent and identically distributed. 

Similar to the literature on tree \citep{xin2022exploring, lin2020generalized} we consider $\bX_{n\times p}$ to be a $n \times p$-dimensional matrix of pretreatment covariates with binary entries where row correspond to different units and columns correspond to different features. Note that you can convert any finite matrix of pre-treatment covariates into a matrix with binary entries \citep{lin2020generalized}.
Now, we describe a pure random tree sampling approach to sample a tree with $Q$ leaves \citep{breiman2000some}:
\begin{enumerate}
    \item At step $0$, we start with a root node
    \item At any $(q-1)$-th step, there are $q$ existing leaf nodes
    \begin{enumerate}
        \item Randomly choose one of them to split and grow the tree
        \item Randomly choose one of the $p$ features to split on
    \end{enumerate}
    \item At $(Q-1)$-th step, the constructed tree has $Q$ leaves
    \begin{enumerate}
        \item For each leave, randomly choose the label from $\{0,1\}$.
    \end{enumerate}
\end{enumerate}
Thus, note that, each tree can be uniquely characterized by the sequence of Q-1 splits $\{o_1, \dots, o_{Q-1}\}$ where $o_i = [\textrm{split-node}_i, \textrm{feature}_i]$ and the labels for each leaf $l_1,\dots,l_Q$ where $\textrm{split-node}_i$ is the node chosen to split and $\textrm{feature}_i$ is the feature to split on in step $i$. Each sampled tree represents a function $w^Q$ such that $w^Q(\bx) = \sum_{i=1}^{Q} \mathbf{1}[\bx \in \textrm{leaf}_i] l_i$. We represent these $w$'s with a superscript $^Q$ to represent that they are characterized by a tree with $Q$ leaves. 

Let $w^Q_* \in \arg \min_{w^Q} \mathcal{L}(w^Q, \setSn)$ denote the optimal tree with $Q$-leaves that minimizes the objective function $\mathcal{L}$. Let $\mathcal{W}^{Q}_K := \{w_1^Q,\dots,w_K^Q\}$ denote a set of $K$ independently and identically sampled trees. Then, we are interested in understanding if $w^Q_* \in \mathcal{W}^{Q}_K$ with high probability. 

For any tree $w^{Q}_k \in \mathcal{W}^{Q}_K$, $$P(w^{Q}_k = w^Q_*) \geq P(\{o^k_{1}, \dots, o^k_{Q-1}\} = \{o^*_{1}, \dots, o^*_{Q-1}\}) P(\{l^k_1,\dots,l^k_Q\} = \{l^*_1,\dots,l^*_Q\}).$$
At q-th step, $P(o^k_q = o^*_q \mid \{o^k_{1}, \dots, o^k_{q-1}\} = \{o^*_{1}, \dots, o^*_{q-1}\}) = \frac{1}{q-1}\frac{1}{p}$. Thus, $P(\{o^k_{1}, \dots, o^k_{Q-1}\} = \{o^*_{1}, \dots, o^*_{Q-1}\}) = \frac{p}{Q-1!}\left(\frac{1}{p}\right)^Q$. Further, $P(\{l^k_1,\dots,l^k_Q\} = \{l^*_1,\dots,l^*_Q\}) = \frac{1}{2^Q}$. Thus $P(w^{Q}_k = w^Q_*) \geq \frac{p}{Q-1!}\left(\frac{1}{2p}\right)^Q$ and $P(w^{Q}_k \neq w^Q_*) \leq 1 - \frac{p}{Q-1!}\left(\frac{1}{2p}\right)^Q.$ Thus, $$P(w^Q_* \notin \mathcal{W}^{Q}_K) \leq \left(1 - \frac{p}{Q-1!}\left(\frac{1}{2p}\right)^Q\right)^K.$$ Thus, as number of sampled trees, $K$, goes to $\infty$, $P(w^Q_* \notin \mathcal{W}^{Q}_K)$ goes to $0$.

Note that these bounds are very loose and for a weaker tree sampling algorithm than ROOT. Further, theoretical work will focus on exploring a tighter bound and a better convergence rate for ROOT. 

%% file: results.tex
\section{Synthetic Data Experiments}\label{sec: synth_exp}
Now, we study the performance of our proposed approach via synthetic data experiments. In particular, we are interested in understanding the extent to which our approach (i) produces interpretable characterizations of underrepresented populations, (ii) reduces the variance of the WTATE relative to the variance of the TATE, and (iii) handles high-dimensional and non-linear data in four data-generative processes (DGPs) with different features.

\subsection{Data Generative Process}\label{sec: dgp}
These four synthetic DGPs are designed to be akin to different real-world scenarios and illustrate important features of our approach. We call them (i) Community, (ii) Box, (iii) Clone, and (iv) High-dimensional DGPs due to their respective properties.

\paragraph{Community DGP.} In our data generation process, units belong to one of two latent communities: Community `A' encompasses 75\% of the population, while Community `B' constitutes the remaining 25\%. The probabilities of units belonging to these communities are defined as follows:
\begin{equation*}
    P( C_i = A ) = \frac{3}{4} \textrm{ and } P( C_i = B ) = \frac{1}{4}
\end{equation*}
For units from community `A', $X_{i,0}$ is drawn from a normal distribution centered around $0$ and $X_{i,1}$ is sampled from a normal distribution centered around $X_{i,0}$. Conversely, for units from community `B', $X_{i,0}$ is drawn from a normal distribution centered around $4$ and the $X_{i,1}$ is drawn from a normal distribution centered around $X_{i,0}$:
\begin{eqnarray*}
    X_{i,0} \overset{iid}{\sim} \mathbbm{1}[ C_i=A ] \mathcal{N}(0,1) + \mathbbm{1}[ C_i=B ] \mathcal{N}(4,1) \textrm{ and }    X_{i,1} \overset{iid}{\sim} \mathcal{N}(X_{i,0},3).
\end{eqnarray*}
The probability of an individual $i$ participating in the trial is determined by the proximity of $(X_{i,0}, X_{i,1})$ to the origin $(0,0)$, with a decreasing probability as the distance from the origin increases. Specifically, $P(S_i = 1 \mid \bX_i) = 0.5 \mathbbm{1}[r_i < 3] + 0.25 \mathbbm{1}[3 \leq r_i < 5]$ where $r := (X_{i,0}^2 + X_{i,1}^2)^{1/2}$. In the trial, the binary treatments are assigned randomly with equal probability: $ T_i \overset{iid}{\sim} \textrm{Bernoulli}\left(0,1/2\right)$. The outcome function is rather simple, in that the potential outcome $Y_i(0)=0$ and the potential outcome $Y_i(1)$ quadratically depend on $X_{i,0}$ and $X_{i,1}$: $Y_i =  T_i \left( (X_{i,0}^2 + X_{i,1}^2) + \epsilon_i \right)$ where $\epsilon_i \sim \mathcal{N}(0,1)$. In summary, this is a low dimensional setup with only two covariates, both of which are effect modifiers as well as selectors. Here, units in community B are underrepresented in the trial data due to the nature of the participation probabilities. Although simple, the non-linearity of the treatment effect as well as the latentness of the communities make this DGP challenging.

\paragraph{Box DGP.} This DGP increases complexity by considering a high-dimensional setting with 100 independent and identically distributed pretreatment covariates: $$X_{i,0}, X_{i,1}, \ldots, X_{i,100} \sim \mathcal{U}(0,1).$$
The probability of selecting any unit $i$ for participation in the trial is uniform and constant at $\text{expit}(1/4)$ across the entire covariate space, except for a specific region defined by covariates: $1/2 < X_{i,0} < 1$ and $1/2 < X_{i,1} < 1$. Within this box-shaped region, the probability of participation decreases significantly:
$$P(S_i=1 \mid \mathbf{X}_i) = \text{expit}\left( \frac{1}{4} - 2\left( \mathbbm{1}\left(\frac{1}{2} < X_{i,0} < 1\right) \times \mathbbm{1}\left(\frac{1}{2} < X_{i,1} < 1\right) \right) \right).$$ 
Treatment assignment is again random with the probability of treatment of 0.5. The potential outcome under control, $Y_i(0)$, is defined by a highly non-linear Friedman's function described in \cite{friedman1991multivariate} and \cite{breiman1996bagging}: $$Y_i(0) = 10 \sin(\pi X_{i,0} X_{i,1}) + 20 (X_{i,2} - 0.5)^2 + 10 X_{i,3} + 5 X_{i,4} + \epsilon_i,$$ where $\epsilon_i \sim \mathcal{N}(0,1)$, and the treatment effect is given by $\log(Y_i(0) + 1).$
 $$Y_i(1) = Y_i(0) + \log(Y_i(0) + 1).$$ Note that, here, covariates $X_{i,0}$ and $X_{i,1}$ are affecting both sample selection $S$ and the treatment effect $Y_i(1) - Y_i(0)$.
In summary, the Box DGP model involves high-dimensional covariates, non-linear response functions, and varying trial participation probabilities based on covariate characteristics. 

\begin{figure}
    \centering
    \includegraphics[width=0.7\textwidth]{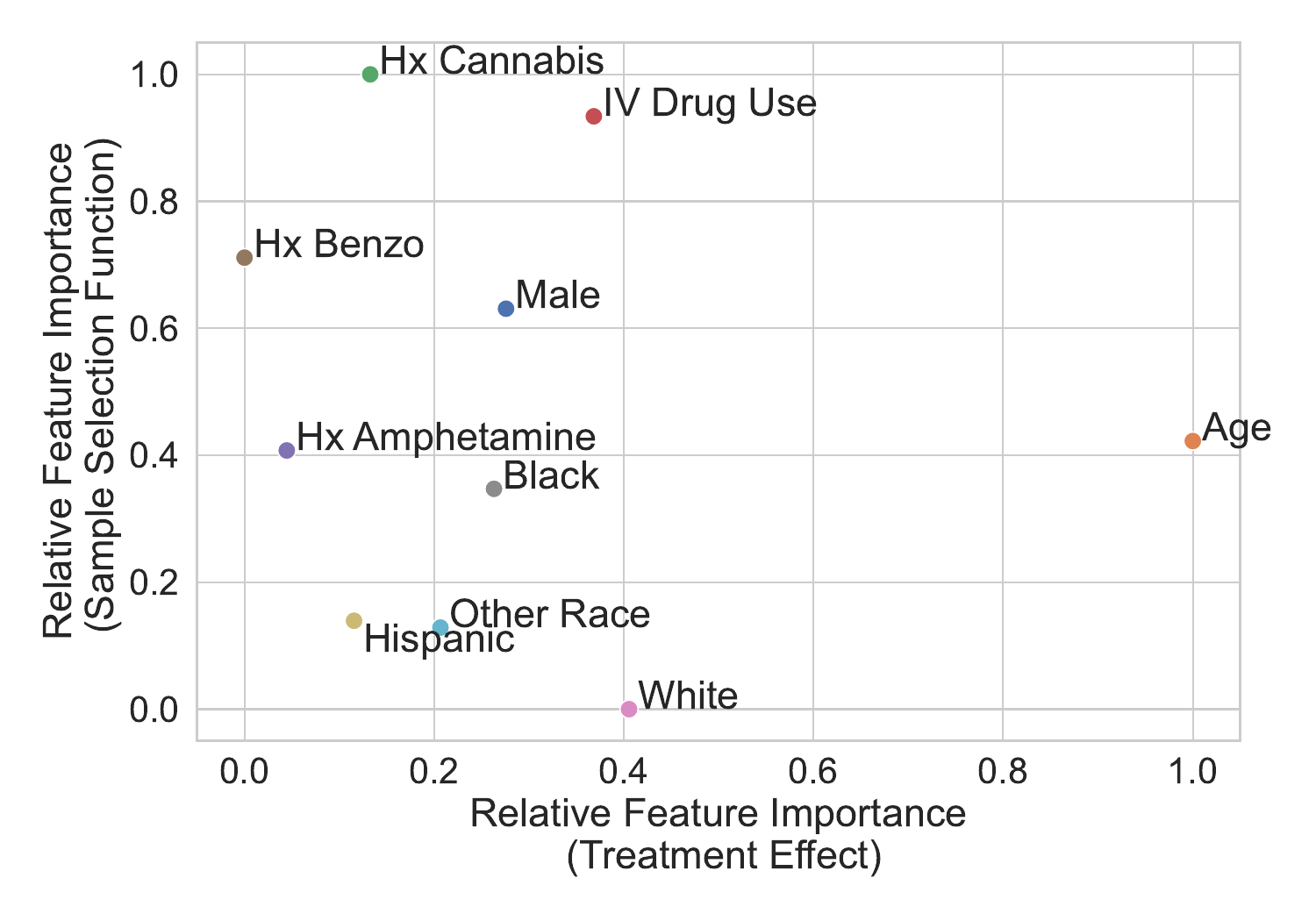}
    \caption{Scatter plot of Clone DGP relative feature importance for sample selection and treatment effect.}
    \label{fig: feature_imp}
\end{figure}

\paragraph{Clone DGP.} We design this DGP to mimic the data distributions in the case study of our interest (presented in Section~\ref{sec: moud}). We provide a detailed description of data in Section~\ref{sec: moud}, however, here, it is important to know that $Y$ is a binary outcome and $\bX$ is a 10-dimensional vector. To generate the synthetic data that mimics our observed data, we need to design the joint distribution $P(Y(1), Y(0),\bX, S, T)$. Leveraging assumption \ref{a: internal_val} and  \ref{a: external_val}, we rewrite the joint as $P(Y(1), Y(0),\bX, S, T) = P(T \mid S, \bX) P(S \mid \bX) P(Y(1),Y(0),\bX)$. We learn the following models using the observed MOUD data: (i) $P( Y(1)=1 \mid \bX, S=1)$ (ii) $P(Y(0)=1 \mid \bX, S=1]$, (iii) $P(T=1 \mid \bX, S=1)$ and (iv) $P(S=1 \mid \bX)$. We use gradient-boosting trees to learn the two potential outcome models and cross-validated logistic regression to learn the propensity and selection scores. We use the potential outcome models to impute the missing potential outcomes in the observed data and then learn the joint distribution $P(Y(1), Y(0), \bX)$. 
And with a slight abuse of notations, to sample a synthetic data with $n$ units, we first sample $\{Y_i(1), Y_i(0), \bX_i\}_{i=1}^n \overset{iid}{\sim} P(Y(1), Y(0), \bX)$. Next, we sample $\{S_i\}$ and then, $\{T_i\}$ for units with $S_i=1$. This data generative procedure guarantees that the empirical distributions of the observed data and the synthetic data are similar to each other. Further, the trained models inform the ground-truth effect modifiers as well as sample selectors for the synthetic data. Figure~\ref{fig: feature_imp} shows the relative importance of each feature in determining the selection probability (y-axis) and treatment effect (x-axis).

\paragraph{High-dimensional DGP.} Similar to the Box DGP, here too we consider a high-dimensional setup with 100 iid covariates per individual: $\{X_{i,0}\dots X_{i,100}\} \overset{iid}{\sim} \mathcal{N}(0,1)$. Unlike the Box DGP, all 100 covariates contribute towards an individual's likelihood of participating in the trial: $P(S_i=1 \mid \bX_i ) = \mathrm{expit}( \alpha_0^T \bX_i )$. The control potential outcome and the treatment effect are linear functions of covariates: $Y_i(0) = \beta_0^T \bX_i + \epsilon_i$, and $Y_i(1) = Y_i(0) + \beta_1^T \bX_i + \epsilon_i$ where $\epsilon_i \sim \mathcal{N}(0,1)$. Here, the parameters $\alpha_0$, $\beta_0$, and $\beta_1$ are also drawn from a standard normal distribution. The analysis challenge in this DGP is that the features affecting the sample selection and treatment effect are high-dimensional. Thus there is no sparse representation of the underrepresented group.

\subsection{Analysis}
We employ our nonparametric ROOT approach and compare it with three alternatives: (i) a single randomized tree (1-tree), (ii) linear parametric modeling of $w$ (Linear-w),  and (iii) modeling $w$ as an indicator function (Indicator w). All four of these approaches are discussed in Section~\ref{sec: method}. Recall that we are directly modeling $w$ as a function of $\bX$ to ensure the interpretability and communicability of the study population. 

We first examine the extent to which our approach reduces the variance of the WTATE estimates compared to the TATE estimates. Table~\ref{tab: synth_exp} shows the standard errors for each of the four approaches as well as the estimates on the unpruned (or unrefined) target. The precision measured as the standard error is a function of study data as well as the estimator. We use the same IPW WTATE estimator across all 4 approaches which is an unbiased estimator of treatment effect. Focusing on the standard errors (precision), we find that for Community, Box, and Clone DGPs, the ROOT approach produces the most precise estimates. For High-dimensional DGP, however, the Linear-w approach tends to outperform other approaches. This is primarily because all the relationships in this DGP are linear and thus $w(\bX) = \gamma^T \bX$ is the ``correct'' specification.



Thus,  our experimental results indicate that our approach improves the precision of the TATE estimate without compromising the bias. Further, given the nonlinearity of the Box DGP, and the high-dimensional nature of both the Box and High-dimensional DGPs, our results also support the ability of our approach to handle a complex and high-dimensional dataset.
\begin{table}[]
    \centering
    \begin{tabular}{@{}l|r|rrrr@{}}
\toprule
\textbf{}        & \begin{tabular}{c}
\textbf{Original} \\ \textbf{Target}\\ \textbf{Estimate}
\end{tabular} & \textbf{ROOT} & \textbf{1-Tree} & \textbf{Linear} & \textbf{Indicator} \\ \midrule
\textbf{Community DGP} &                   &               &                 &                 &                    \\
\quad \textit{Std. Error}    & 3.073             & \textbf{0.449}         & 0.985           & 2.995           & 0.770              \\ \midrule
\textbf{Box DGP}       &                   &               &                 &                 &                    \\
\quad \textit{Std. Error}    & 0.910             & \textbf{0.324}         & 0.409           & 0.910           & 0.763              \\ \midrule
\textbf{Clone DGP}    &                   &               &                 &                 &                    \\

\quad \textit{Std. Error}    & 4.655             & \textbf{4.520}        & 4.655           &  4.655        & 4.551 \\ \midrule
\textbf{High-dim. DGP}    &                   &               &                 &                 &                    \\

\quad \textit{Std. Error}    & 0.191             & 0.162        & 0.169           & \textbf{0.127}           & 0.170              \\ \bottomrule
\end{tabular}
    \caption{\textit{Table of Synthetic Data Analysis Results.} Precision (measured as standard error) for the four DGPs described in Section~\ref{sec: dgp}, and four approaches described in Section~\ref{sec: method}. ROOT is our proposed approach while the other three are baseline comparison methods.}
    \label{tab: synth_exp}
\end{table}

\begin{figure}
    \centering
    \begin{tabular}{c}
         \includegraphics[width = 0.6\textwidth]{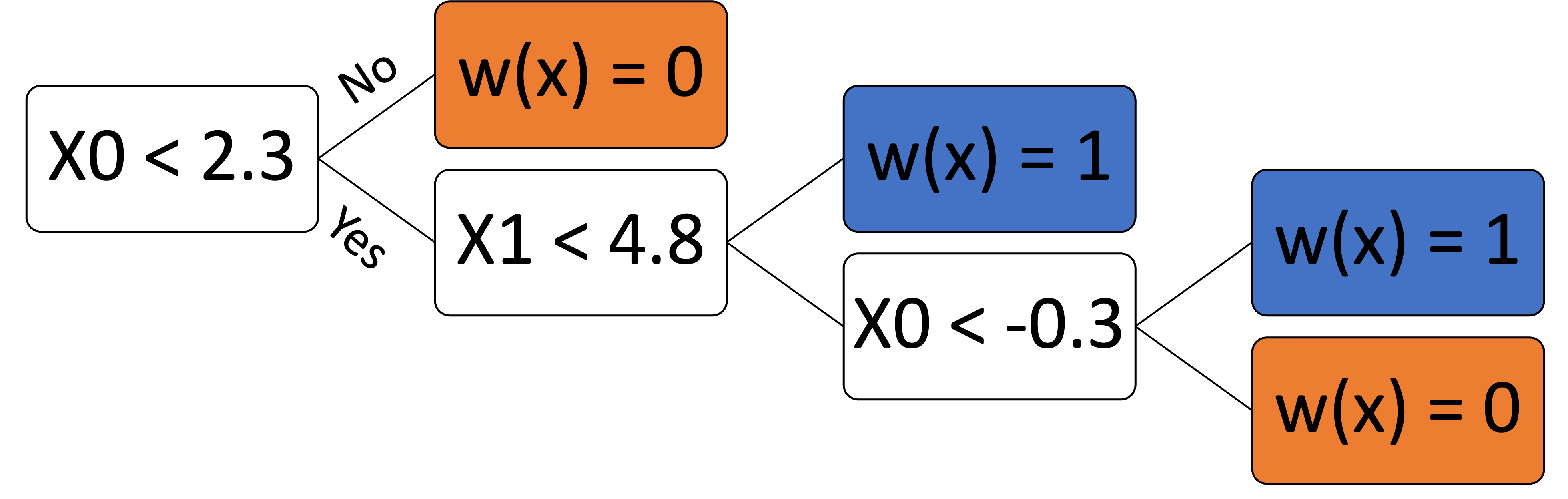} \\ (a) Community DGP \\ \includegraphics[width = 0.5\textwidth]{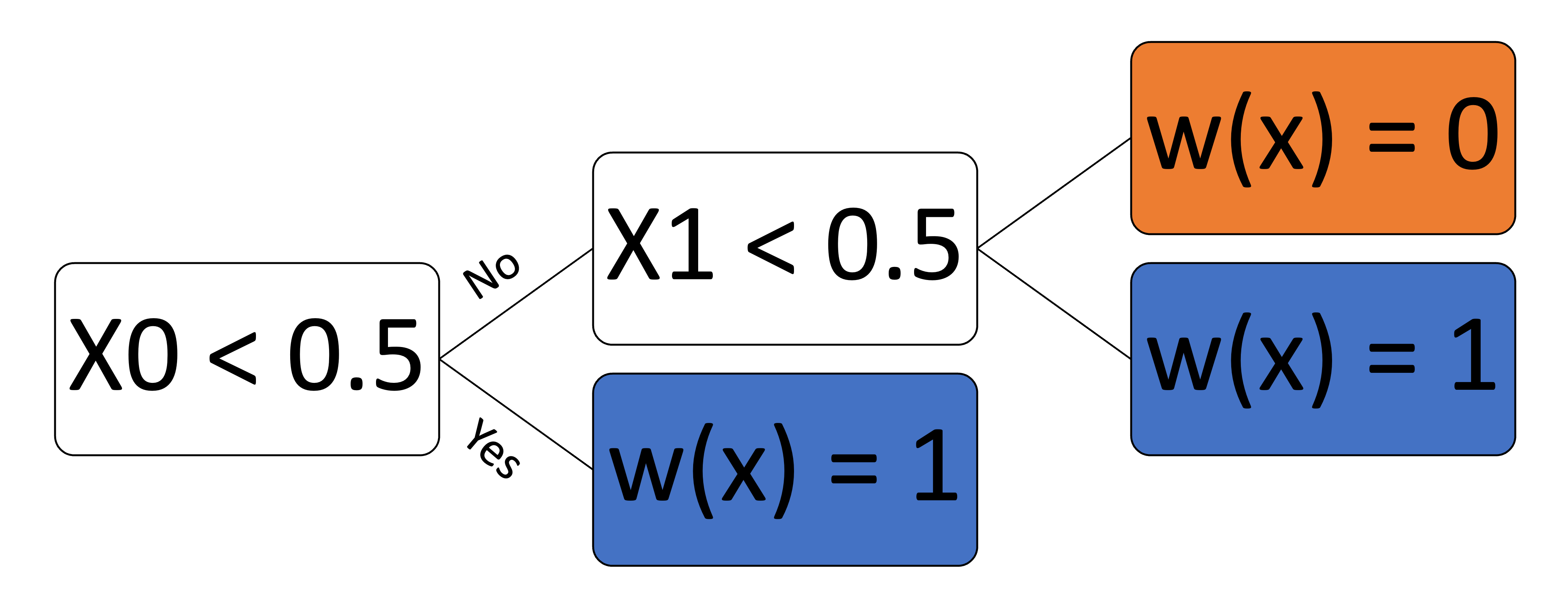} \\ (b) Box DGP \\
         \\ \includegraphics[width = 0.6\textwidth]{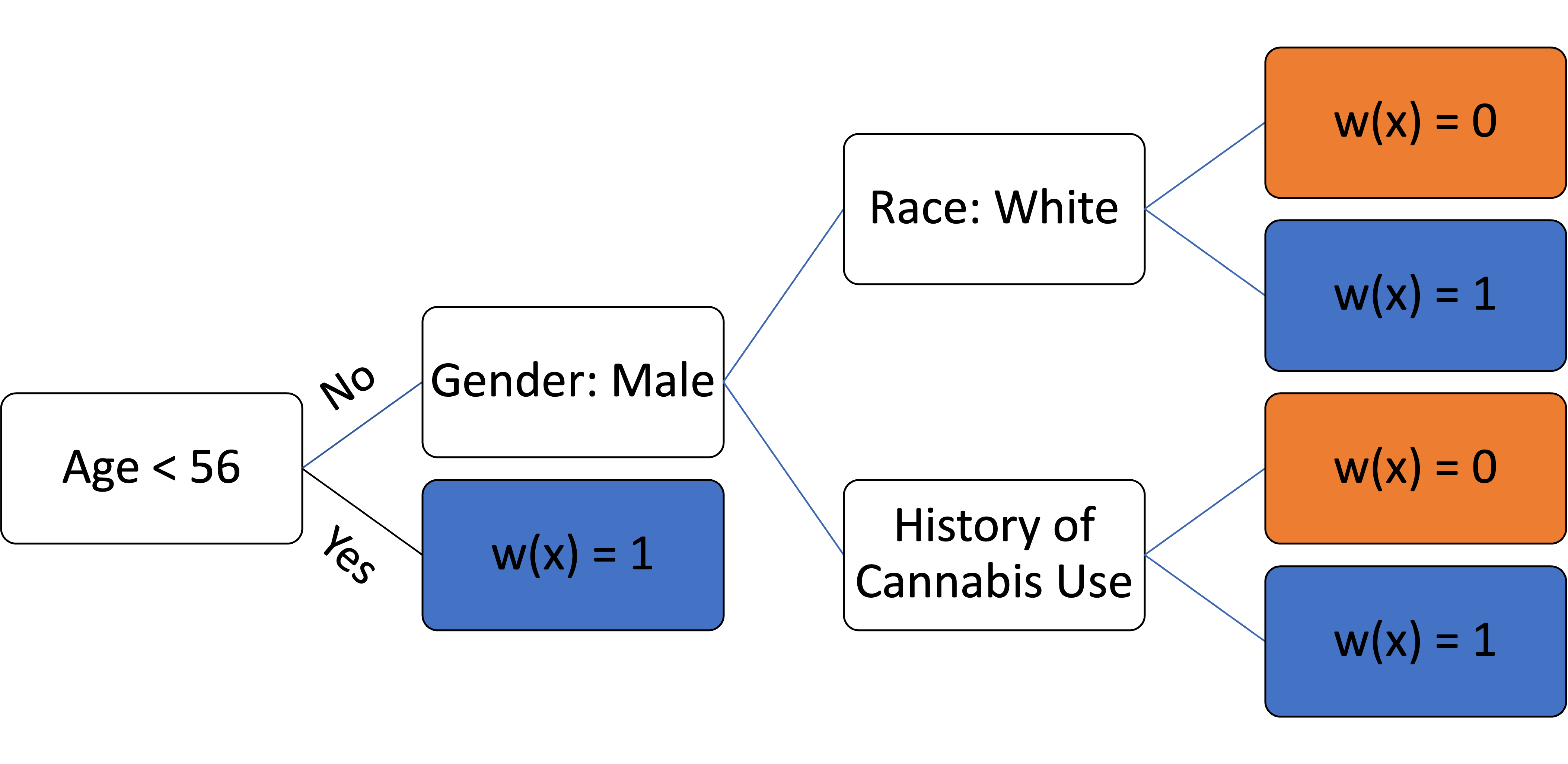} \\ (c) Clone DGP \\
         \includegraphics[width = 0.6\textwidth]{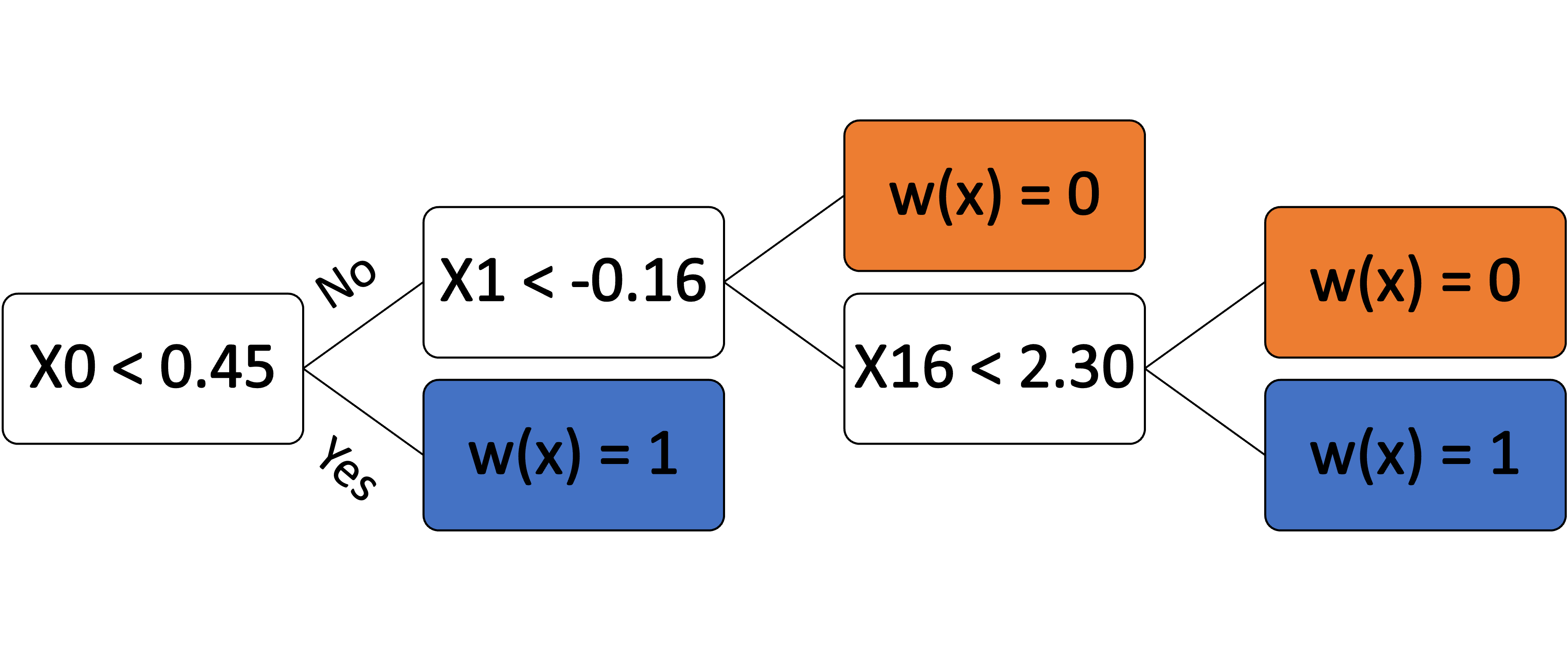} \\
          (d) High-dim. DGP
    \end{tabular}
    \caption{Tree characterizing indicating the study population in Blue and the underrepresented subgroups in Orange for all four DGPs}
    \label{fig: synth_tree}
\end{figure}

\begin{figure}
    \centering
    \begin{tabular}{ccc}
        \includegraphics[width = 0.3\textwidth]{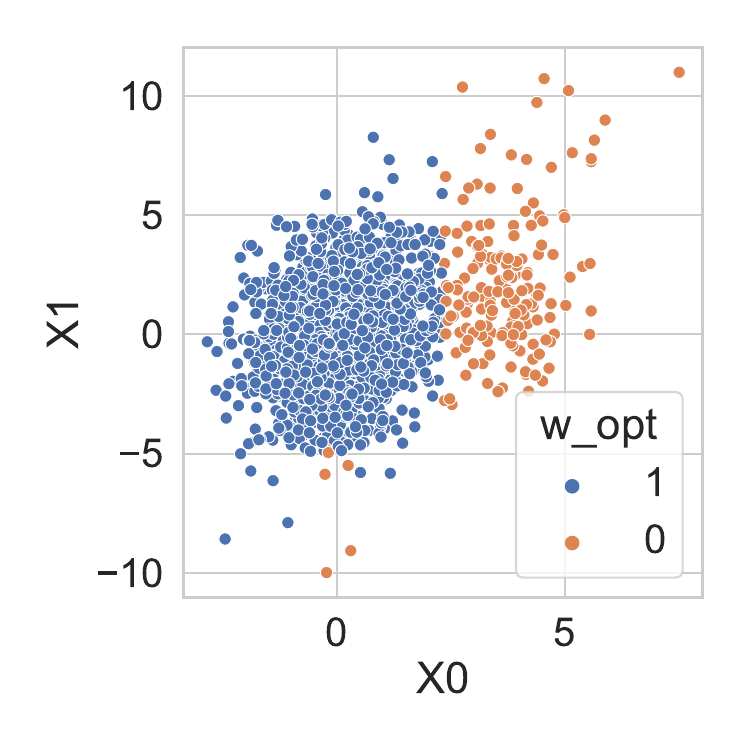} & 
        \includegraphics[width = 0.3\textwidth]{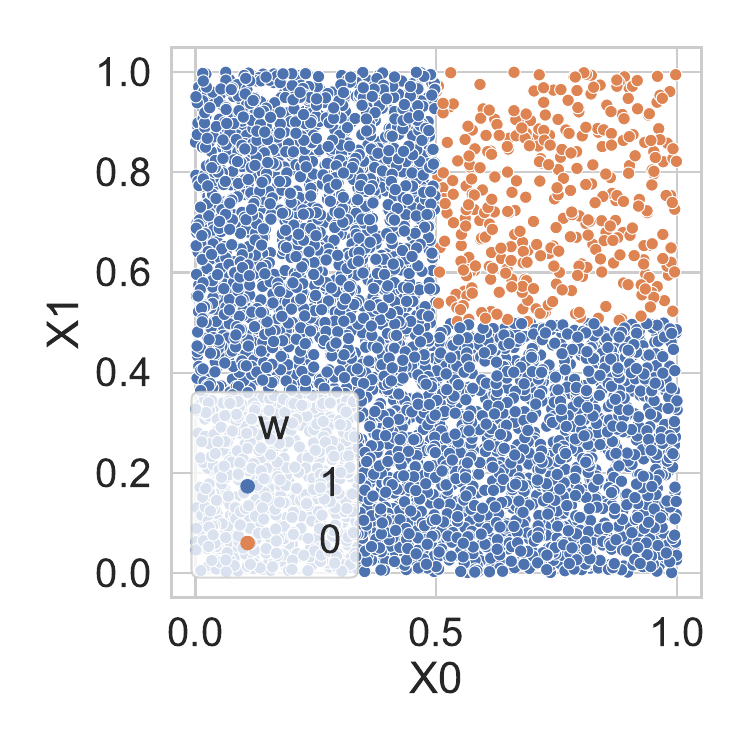} & 
        \includegraphics[width = 0.3\textwidth]{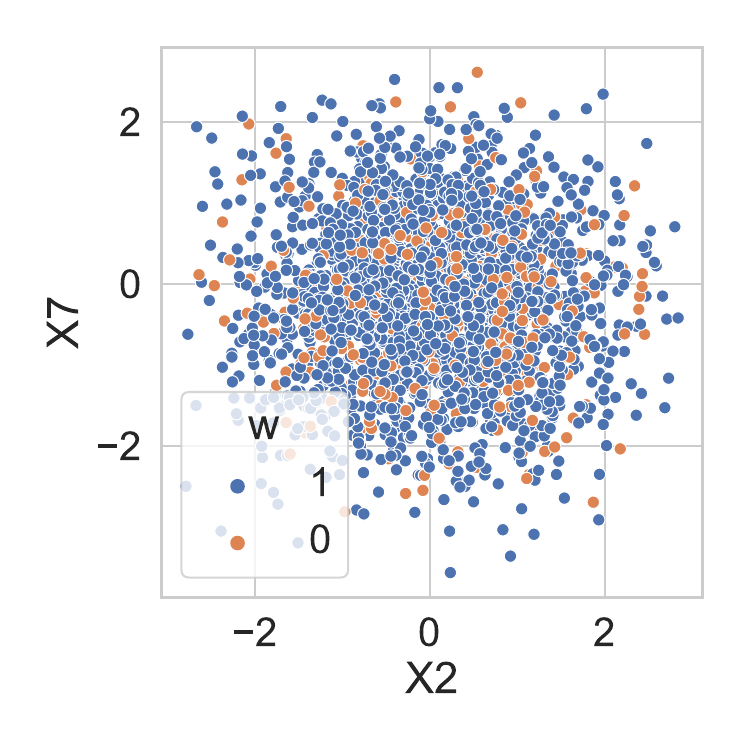} \\
        (a) Community DGP & (b) Box DGP & (c) High-dim. DGP
    \end{tabular}
    \caption{Scatter plots projecting the data distribution across the two most important covariates in the true DGP. The study populations are indicated in Blue and the underrepresented subgroups are in Orange for all three DGPs}
    \label{fig: synth_two_dim}
\end{figure}

Next, we discuss how one can use the ROOT approach to produce interpretable characteristics of the underrepresented groups. In Figure~\ref{fig: synth_tree}, we shows trees characterizing the insufficiently represented populations for the four DGPs. Further, Figure~\ref{fig: synth_two_dim} shows the data distribution color-coded based on the predicted values of $w(\bX)$ using ROOT for each unit in the dataset across the two most important features for each of the DGPs that are both outcome modifiers as well as the sample selectors. Our results shown in these figures indicate that ROOT is not only correctly able to identify the underrepresented groups or communities but also produces a tree that can be used to communicate their characteristics easily. 

Unlike the results for the Community, Box, and Clone DGPs, in the case of the High-dimensional DGP, the decision boundary in the two-dimensional projection is not sharp. This is because in this scenario all of the 100 covariates are moderators and relate to trial participation and no two-dimensional slicing of the covariate space can perfectly show the decision boundary. 

Further, for the high-dimensional DGP, it is worth noting that ROOT is only the second-best approach in terms of precision even this DGP is designed to be challenging for ROOT. This is because the default hyperparameters of ROOT are chosen to find subgroups with sparse representations (i.e., the ones described via shallow trees) for the sake of interpretability. Due to a large number of moderators being almost equally important in the construction of this DGP, the shallow representation yields a less-than-optimal representation of the underrepresented population. This is because of the large number of moderators being almost equally important and thus there is no shallow tree that can accurately characterize the underrepresented groups.

%% file: appendix_table.tex
\section{MOUD Data Descriptive Statistics}
We consider methadone as $T=1$ and buprenorphine as $T=0$. For the outcome, $Y=1$ indicates relapse by 24 weeks after treatment assignment. We only consider the pre-treatment covariates observed in both the trial and target data. Our set of pre-treatment covariates includes the participants' age, race and biological sex along with substance use history (amphetamine, benzodiazepines, cannabis and IV drug use) assessed at the initiation of MOUD treatment. We provide Table~\ref{tab: data_moud} for a listing of all of the covariates included.
\input{table_data}

%% file: table_data.tex
\begin{table}
    \centering
    \begin{tabular}{lcc}
        \toprule
        {} & \textbf{TEDS-A} & \textbf{START} \\
        {} &  S=0 & S=1 \\
        \midrule
        Sample Size & 740,015 & 1,271\\
        \midrule
        \textit{Covariates} & & \\
        \quad Male (\%)         &   58\% &   68\% \\
        \quad Age (IQR)            &  [27y,47y] &  [27y,46y] \\
        \quad History of Benzo Use (\%)     &   6\% &   17\% \\
        \quad History of Amphetamine Use (\%) &   7\% &   13\% \\
        \quad History of IV Drug Use (\%)   &   54\% &   70\% \\
        \quad History of Cannabis Use (\%)   &   11\% &   29\% \\
        \quad Race & & \\
        \quad \quad White (\%)       &   65\% &   68\% \\
        \quad \quad Black (\%)         &   15\% &   9\% \\
        \quad \quad Hispanic (\%)       &   14\% &   16\% \\
        \quad \quad Other Race (\%)     &   5\% &   8\% \\
        \midrule
        \textit{Treatment}         &   &   \\
        \quad Methadone (\%) & - & 42\%\\
        \quad Buprenorphine (\%) & - & 58\%\\
        \textit{Outcome}    &  &    \\
        \quad Relapse in 24w (\%) & - & 71\% \\
        \bottomrule
    \end{tabular}
    \caption{The distributions of demographic and clinical pre-treatment covariates, treatments, and outcomes across the START trial and TEDS-A datasets.}
    \label{tab: data_moud}
\end{table}

%% file: moud_discuss_assume.tex
\section{Discussion on Assumptions for MOUD Case Study}
\label{sec: discuss_assume} 
\begin{figure}
    \centering
    \includegraphics[width=0.6\textwidth]{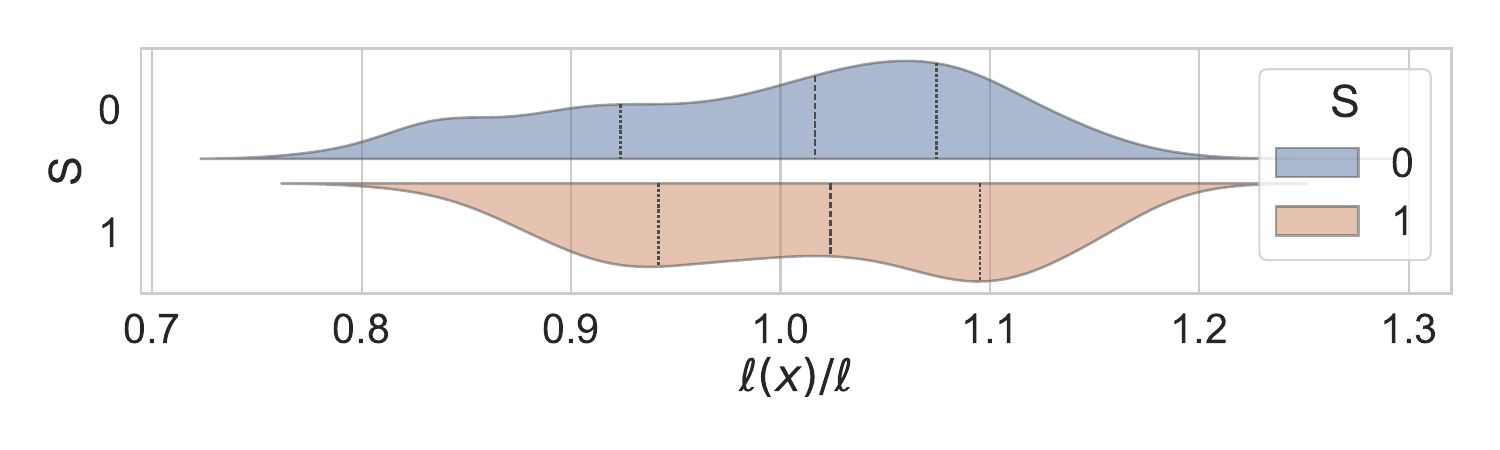}
    \caption{Violin-plot showing the distribution of the normalized selection odds $\ell(\bx)/\ell$ for the units in trial and target samples.}
    \label{fig: positive}
\end{figure}
    \begin{itemize}
        \item (A1) This assumption can be assessed through examination of the data; Figure~4 depicts the overlap between the distributions of the target population and trial cohort on normalized conditional selection odds (i.e., $\frac{\ell(x)}{\ell} = \frac{1-\pi}{\pi} \frac{\pi(x)}{1-\pi(x)}$). We observe that there is significant overlap of the estimated selection odds between the trial and target samples. Further, none of the units have selection probabilities and thus selection odds equal to 0. Thus, structural positivity assumption (A1) appears to hold in this data, i.e., $P(S=1 \mid X=x) \neq 0$ for all units in the target sample. 
        \item (A2) In our study, we focus on transporting the intent-to-treat effect using the START trial data. Given the randomized design of the trial  \cite{saxon2013buprenorphine}, unconfounded treatment assignment (A2) holds.
         \item (A3) A3 assumes the absence of an unobserved factor, $U$, affecting both the outcome as well as the trial selection probability once the set of observed covariates $x$ are adjusted for. This assumption is untestable. Based on domain knowledge, accounting for $x$ may be sufficient to ensure the exchangeability of the potential outcomes across $S=0$ and $S=1$ \citep{rudolph2021optimizing, rudolph2022under, nunes2020moderators, rudolph2023optimally}. However, certain potentially important variables such as homelessness, parole and probation status, medication preference are unobserved across datasets and could potentially lead to violation of A3 \citep{nunes2020moderators}.
         \item (A4) Stable unit treatment value assumption (SUTVA) has two parts: (i) there is no interference and (ii) there is a single version of each of the treatments across all units. The first part may be violated because 1) individuals are enrolled across 8 opioid treatment programs, and 2) within each treatment program, individuals may be part of common social networks. The same limitations regarding patients nested within program and with overlapping social networks exist in the TEDS-A data. How to best address these complex dependencies is an area for future work.  For now we assume any such SUTVA violation is negligible. Further, the second part may be challenged because the MOUD doses are not standardized across treatment programs. However, our work focuses on estimating intent to treatment effect and thus, we assume that this part of SUTVA holds as well.
    \end{itemize}

%% file: main.bbl
\begin{thebibliography}{}

\bibitem[Abuse et~al., 2020]{abuse2020treatment}
Abuse, S. et~al. (2020).
\newblock Treatment episode data set admissions (teds-a).

\bibitem[Breiman, 1996]{breiman1996bagging}
Breiman, L. (1996).
\newblock Bagging predictors.
\newblock {\em Machine learning}, 24:123--140.

\bibitem[Breiman, 2000]{breiman2000some}
Breiman, L. (2000).
\newblock Some infinity theory for predictor ensembles.
\newblock Technical report, Citeseer.

\bibitem[Breiman, 2001]{breiman2001statistical}
Breiman, L. (2001).
\newblock Statistical modeling: The two cultures (with comments and a rejoinder by the author).
\newblock {\em Statistical science}, 16(3):199--231.

\bibitem[Breiman, 2017]{breiman2017classification}
Breiman, L. (2017).
\newblock {\em Classification and regression trees}.
\newblock Routledge.

\bibitem[Burchett et~al., 2011]{burchett2011}
Burchett, H., Umoquit, M., and Dobrow, M. (2011).
\newblock How do we know when research from one setting can be useful in another? {{A}} review of external validity, applicability and transferability frameworks.
\newblock {\em Journal of health services research \& policy}, 16(4):238--244.

\bibitem[Cahan et~al., 2017]{cahan2017computer}
Cahan, A., Cahan, S., and Cimino, J.~J. (2017).
\newblock Computer-aided assessment of the generalizability of clinical trial results.
\newblock {\em International journal of medical informatics}, 99:60--66.

\bibitem[Chipman et~al., 2010]{chipman2010bart}
Chipman, H.~A., George, E.~I., and McCulloch, R.~E. (2010).
\newblock Bart: Bayesian additive regression trees.

\bibitem[Colnet et~al., 2021]{colnet2021generalizing}
Colnet, B., Josse, J., Scornet, E., and Varoquaux, G. (2021).
\newblock Generalizing a causal effect: sensitivity analysis and missing covariates.
\newblock {\em arXiv preprint arXiv:2105.06435}.

\bibitem[Crump et~al., 2009]{crump2009dealing}
Crump, R.~K., Hotz, V.~J., Imbens, G.~W., and Mitnik, O.~A. (2009).
\newblock Dealing with limited overlap in estimation of average treatment effects.
\newblock {\em Biometrika}, 96(1):187--199.

\bibitem[Dahabreh et~al., 2022]{dahabreh2022global}
Dahabreh, I.~J., Robins, J.~M., Haneuse, S.~J., Robertson, S.~E., Steingrimsson, J.~A., and Hern{\'a}n, M.~A. (2022).
\newblock Global sensitivity analysis for studies extending inferences from a randomized trial to a target population.
\newblock {\em arXiv preprint arXiv:2207.09982}.

\bibitem[Degtiar and Rose, 2023]{degtiar2023review}
Degtiar, I. and Rose, S. (2023).
\newblock A review of generalizability and transportability.
\newblock {\em Annual Review of Statistics and Its Application}, 10:501--524.

\bibitem[Dekkers et~al., 2010]{dekkers2010}
Dekkers, O.~M., {von Elm}, E., Algra, A., Romijn, J.~A., and Vandenbroucke, J.~P. (2010).
\newblock How to assess the external validity of therapeutic trials: A conceptual approach.
\newblock {\em International Journal of Epidemiology}, 39:89--94.

\bibitem[Ding, 2023]{ding2023first}
Ding, P. (2023).
\newblock A first course in causal inference.
\newblock {\em arXiv preprint arXiv:2305.18793}.

\bibitem[Ding et~al., 2016]{ding2016}
Ding, P., Feller, A., and Miratrix, L. (2016).
\newblock Randomization inference for treatment effect variation.
\newblock {\em Journal of the Royal Statistical Society. Series B, Statistical methodology}, 78(3):655--671.

\bibitem[Egami and Hartman, 2023]{egami2023elements}
Egami, N. and Hartman, E. (2023).
\newblock Elements of external validity: Framework, design, and analysis.
\newblock {\em American Political Science Review}, 117(3):1070--1088.

\bibitem[Friedman, 1991]{friedman1991multivariate}
Friedman, J.~H. (1991).
\newblock Multivariate adaptive regression splines.
\newblock {\em The annals of statistics}, 19(1):1--67.

\bibitem[Green and Glasgow, 2006]{green2006}
Green, L.~W. and Glasgow, R.~E. (2006).
\newblock Evaluating the relevance, generalization, and applicability of research: Issues in external validation and translation methodology.
\newblock {\em Evaluation \& the health professions}, 29:126--153.

\bibitem[Greenhouse et~al., 2008a]{greenhouse2008}
Greenhouse, Kelleher, K., Seltman, H., and Gardner, W. (2008a).
\newblock Generalizing from clinical trial data: A case study. the risk of suicidality among pediatric antidepressant users.
\newblock {\em Statistics in Medicine}, 27(11):1801--13.

\bibitem[Greenhouse et~al., 2008b]{greenhouse2008generalizing}
Greenhouse, J.~B., Kaizar, E.~E., Kelleher, K., Seltman, H., and Gardner, W. (2008b).
\newblock Generalizing from clinical trial data: a case study. the risk of suicidality among pediatric antidepressant users.
\newblock {\em Statistics in medicine}, 27(11):1801--1813.

\bibitem[Hser et~al., 2014]{hser2014treatment}
Hser, Y.-I., Saxon, A.~J., Huang, D., Hasson, A., Thomas, C., Hillhouse, M., Jacobs, P., Teruya, C., McLaughlin, P., Wiest, K., et~al. (2014).
\newblock Treatment retention among patients randomized to buprenorphine/naloxone compared to methadone in a multi-site trial.
\newblock {\em Addiction}, 109(1):79--87.

\bibitem[Huang, 2024]{huang2024sensitivity}
Huang, M.~Y. (2024).
\newblock Sensitivity analysis for the generalization of experimental results.
\newblock {\em Journal of the Royal Statistical Society Series A: Statistics in Society}, page qnae012.

\bibitem[Iyer et~al., 2023]{iyer2023wide}
Iyer, N., Thejas, V., Kwatra, N., Ramjee, R., and Sivathanu, M. (2023).
\newblock Wide-minima density hypothesis and the explore-exploit learning rate schedule.
\newblock {\em Journal of Machine Learning Research}, 24(65):1--37.

\bibitem[Li et~al., 2019]{li2019addressing}
Li, F., Thomas, L.~E., and Li, F. (2019).
\newblock Addressing extreme propensity scores via the overlap weights.
\newblock {\em American journal of epidemiology}, 188(1):250--257.

\bibitem[Lin et~al., 2020]{lin2020generalized}
Lin, J., Zhong, C., Hu, D., Rudin, C., and Seltzer, M. (2020).
\newblock Generalized and scalable optimal sparse decision trees.
\newblock In {\em International Conference on Machine Learning}, pages 6150--6160. PMLR.

\bibitem[Nguyen et~al., 2017]{nguyen2017sensitivity}
Nguyen, T.~Q., Ebnesajjad, C., Cole, S.~R., and Stuart, E.~A. (2017).
\newblock Sensitivity analysis for an unobserved moderator in rct-to-target-population generalization of treatment effects.
\newblock {\em The Annals of Applied Statistics}, pages 225--247.

\bibitem[Nie et~al., 2021]{nie2021covariate}
Nie, X., Imbens, G., and Wager, S. (2021).
\newblock Covariate balancing sensitivity analysis for extrapolating randomized trials across locations.
\newblock {\em arXiv preprint arXiv:2112.04723}.

\bibitem[Nunes~Jr et~al., 2021]{nunes2020moderators}
Nunes~Jr, E.~V., Scodes, J.~M., Pavlicova, M., Lee, J.~D., Novo, P., Campbell, A.~N., and Rotrosen, J. (2021).
\newblock Sublingual buprenorphine-naloxone compared with injection naltrexone for opioid use disorder: Potential utility of patient characteristics in guiding choice of treatment.
\newblock {\em American Journal of Psychiatry}, pages appi--ajp.

\bibitem[Pearl and Bareinboim, 2011]{pearl2011}
Pearl, J. and Bareinboim, E. (2011).
\newblock Transportability of causal and statistical relations: A formal approach.
\newblock In {\em 2011 {{IEEE}} 11th {{International Conference}} on {{Data Mining Workshops}}}, pages 540--547, {Vancouver, BC, Canada}. {IEEE}.

\bibitem[Petersen et~al., 2012]{petersen2012diagnosing}
Petersen, M.~L., Porter, K.~E., Gruber, S., Wang, Y., and Van Der~Laan, M.~J. (2012).
\newblock Diagnosing and responding to violations in the positivity assumption.
\newblock {\em Statistical methods in medical research}, 21(1):31--54.

\bibitem[Rothwell, 2005]{rothwell2005}
Rothwell, P.~M. (2005).
\newblock External validity of randomised controlled trials: ``to whom do the results of this trial apply?''.
\newblock {\em The Lancet}, 365(9453):82--93.

\bibitem[Rudolph et~al., 2021]{rudolph2021optimizing}
Rudolph, K.~E., D{\'\i}az, I., Luo, S.~X., Rotrosen, J., and Nunes, E.~V. (2021).
\newblock Optimizing opioid use disorder treatment with naltrexone or buprenorphine.
\newblock {\em Drug and alcohol dependence}, 228:109031.

\bibitem[Rudolph and Laan, 2017]{rudolph2017robust}
Rudolph, K.~E. and Laan, M.~J. (2017).
\newblock Robust estimation of encouragement design intervention effects transported across sites.
\newblock {\em Journal of the Royal Statistical Society Series B: Statistical Methodology}, 79(5):1509--1525.

\bibitem[Rudolph et~al., 2022]{rudolph2022under}
Rudolph, K.~E., Russell, M., Luo, S.~X., Rotrosen, J., and Nunes, E.~V. (2022).
\newblock Under-representation of key demographic groups in opioid use disorder trials.
\newblock {\em Drug and alcohol dependence reports}, 4:100084.

\bibitem[Rudolph et~al., 2023a]{rudolph2023optimally}
Rudolph, K.~E., Williams, N.~T., D{\'\i}az, I., Luo, S.~X., Rotrosen, J., and Nunes, E.~V. (2023a).
\newblock Optimally choosing medication type for patients with opioid use disorder.
\newblock {\em American Journal of Epidemiology}, 192(5):748--756.

\bibitem[Rudolph et~al., 2023b]{rudolph2023efficiently}
Rudolph, K.~E., Williams, N.~T., Stuart, E.~A., and Diaz, I. (2023b).
\newblock Efficiently transporting average treatment effects using a sufficient subset of effect modifiers.
\newblock {\em arXiv preprint arXiv:2304.00117}.

\bibitem[Ruggieri, 2019]{ruggieri2019complete}
Ruggieri, S. (2019).
\newblock Complete search for feature selection in decision trees.
\newblock {\em Journal of Machine Learning Research}, 20(104):1--34.

\bibitem[Saxon et~al., 2013]{saxon2013buprenorphine}
Saxon, A.~J., Ling, W., Hillhouse, M., Thomas, C., Hasson, A., Ang, A., Doraimani, G., Tasissa, G., Lokhnygina, Y., Leimberger, J., et~al. (2013).
\newblock Buprenorphine/naloxone and methadone effects on laboratory indices of liver health: a randomized trial.
\newblock {\em Drug and alcohol dependence}, 128(1-2):71--76.

\bibitem[Semenova et~al., 2022]{semenova2022existence}
Semenova, L., Rudin, C., and Parr, R. (2022).
\newblock On the existence of simpler machine learning models.
\newblock In {\em Proceedings of the 2022 ACM Conference on Fairness, Accountability, and Transparency}, pages 1827--1858.

\bibitem[Stuart et~al., 2011]{stuart2011}
Stuart, E.~A., Cole, S.~R., Bradshaw, C.~P., and Leaf, P.~J. (2011).
\newblock The use of propensity scores to assess the generalizability of results from randomized trials: Use of propensity scores to assess generalizability.
\newblock {\em Journal of the Royal Statistical Society: Series A (Statistics in Society)}, 174(2):369--386.

\bibitem[St{\"u}rmer et~al., 2021]{sturmer2021propensity}
St{\"u}rmer, T., Webster-Clark, M., Lund, J.~L., Wyss, R., Ellis, A.~R., Lunt, M., Rothman, K.~J., and Glynn, R.~J. (2021).
\newblock Propensity score weighting and trimming strategies for reducing variance and bias of treatment effect estimates: a simulation study.
\newblock {\em American journal of epidemiology}, 190(8):1659--1670.

\bibitem[Tai et~al., 2011]{tai2011national}
Tai, B., Sparenborg, S., Liu, D., and Straus, M. (2011).
\newblock The national drug abuse treatment clinical trials network: Forging a partnership between research knowledge and community practice.
\newblock {\em Substance Abuse and Rehabilitation}, 2:21.

\bibitem[Tipton, 2013]{tipton2013}
Tipton, E. (2013).
\newblock Improving generalizations from experiments using propensity score subclassification: Assumptions, properties, and contexts.
\newblock {\em Journal of Educational and Behavioral Statistics}, 38(3):239--266.

\bibitem[Tipton, 2014]{tipton2014}
Tipton, E. (2014).
\newblock How generalizable is your experiment? {{An}} index for comparing experimental samples and populations.
\newblock {\em Journal of Educational and Behavioral Statistics}, 39(6):478--501.

\bibitem[Tipton and Peck, 2017]{tipton2017}
Tipton, E. and Peck, L.~R. (2017).
\newblock A design-based approach to improve external validity in welfare policy evaluations.
\newblock {\em Evaluation Review}, 41(4):326--356.

\bibitem[Xin et~al., 2022]{xin2022exploring}
Xin, R., Zhong, C., Chen, Z., Takagi, T., Seltzer, M., and Rudin, C. (2022).
\newblock Exploring the whole rashomon set of sparse decision trees.
\newblock {\em Advances in neural information processing systems}, 35:14071--14084.

\bibitem[Zeng et~al., 2023]{zeng2023efficient}
Zeng, Z., Kennedy, E.~H., Bodnar, L.~M., and Naimi, A.~I. (2023).
\newblock Efficient generalization and transportation.
\newblock {\em arXiv preprint arXiv:2302.00092}.

\end{thebibliography}
